\documentclass[secnumarabic,amssymb, nobibnotes, aps, prd, 11pt, abstract]{revtex4-2}
\usepackage{tocloft}
\usepackage[dvips]{graphicx}
\usepackage{amsmath}
\usepackage{subcaption}
\usepackage{hyperref}
\usepackage{times}
\usepackage{filecontents}
\hypersetup{colorlinks,linkcolor={red},citecolor={blue},urlcolor={red}}
\usepackage{setspace}
\usepackage{xcolor}
\pagecolor{white}

\usepackage{multirow}
\color{black}

\begin{document}

\singlespacing

	\title{Probing Thermodynamic Phase Transitions of 4D \textit{R}-Charged Black Holes via Lyapunov Exponent}%

	\author{Mozib Bin Awal$^1$}
	
	\email{$mozibawal@gmail.com$}
	
	\author{Prabwal Phukon$^{1,2}$}
	\email{$prabwal@dibru.ac.in$}
	
	\affiliation{$^1$Department of Physics, Dibrugarh University, Dibrugarh, Assam,786004.\\$^2$Theoretical Physics Division, Centre for Atmospheric Studies, Dibrugarh University, Dibrugarh,Assam,786004.\\}

	\begin{abstract}
	
We investigate the thermodynamic phase structure of four-dimensional \textit{R}-charged black holes--characterized by four independent $U(1)$ charges--through the lens of Lyapunov exponents associated with unstable circular orbits of both massless and massive particles. Considering three distinct charge configurations (equal, partially unequal, and fully unequal), we show that the thermal profile of the Lyapunov exponent can effectively probe the black hole phase transitions. Furthermore, we demonstrate that the discontinuous jump in the Lyapunov exponent acts as an order parameter, with the associated critical exponent $\delta = 1/2$ near the critical point--consistent across all charge configurations studied.  This supports the proposed universality of the exponent and the broader connection between dynamical chaos and thermodynamic phase transitions.
 
\end{abstract}
	
	\maketitle
	
\section{Introduction}\label{sec0}

Ever since the foundational contributions of Bekenstein, Hawking and others, black hole thermodynamics \cite{Phys,bekens,Hawking,Hawking2,Bardeen} has developed into one of the most fascinating and active areas of research in modern physics. It has emerged as a very powerful framework for probing the deeper connection between general relativity, quantum field theory, thermodynamics and statistical mechanics. Since it is well established that black holes behave as thermodynamic objects, it is natural to ask whether they also undergo processes analogous to those in conventional thermodynamic systems. For instance, initially it was P.C.W Davies and P Hut who studied the ``Phase Transitions" in black holes \cite{Davies}. With Maldacena's $1997$ proposal of the AdS/CFT correspondence \cite{Maldacena}, Anti de Sitter (AdS) black holes in particular have attracted significant attention. Following the inclusion of the cosmological constant as a thermodynamic pressure, numerous studies have investigated the resulting enriched phase structure, revealing striking similarities between black hole thermodynamics and fluid systems, particularly in the context of phase transitions \cite{Kubiz,Hawkpage,Cai,Kastor,Dolan,Dolan2,Dolan3,Kubizna,Xu,Xu2,Zhang}. The standard approach to studying the phase transition in black holes is by analyzing the free energies. However, in the recent times, alternative approach of studying phase transitions in black holes has been gaining more popularity. One such approach is by studying the geometry of the thermodynamic space, particularly using the Ruppeiner geometry \cite{Ruppeiner:2012uc,Miao:2017cyt,Guo:2019oad,Wei:2019yvs,Wang:2019cax,Yerra:2020oph,Yerra:2021hnh}. Another approach that gained popularity lately is by studying the thermodynamic topology \cite{Wu:2022whe,Liu:2022aqt,Fan:2022bsq,Gogoi:2023xzy,Ali:2023zww,Saleem:2023oue,Shahzad:2023cis,Chen:2023elp,Bai:2022klw,Yerra:2022alz,Hazarika:2023iwp} etc. In addition to theoretical investigations, several noteworthy studies have aimed to establish connections between the thermodynamic phase transitions of black holes and potentially observable astrophysical signatures. These include quasinormal modes (QNMs)\cite{Liu:2014gvf,Zou:2017juz,Zhang:2020khz,Mahapatra:2016dae,Chabab:2016cem}, which characterize the damped oscillations of perturbations around black holes, circular orbit radius of test particles \cite{Wei:2017mwc,Wei:2018aqm,Zhang:2019tzi} and the shadow radius of black holes \cite{Zhang:2019glo,Belhaj:2020nqy}.

For investigating the dynamical behaviour of systems that exhibits extreme sensitivity to initial conditions, chaos theory, a field that emerged in physics and mathematics is used. At the core of chaos theory lies the concept of the Lyapunov exponent \cite{lyp}, which quantifies the rate at which nearby trajectories in phase space diverge or converge over time \cite{lyp2}. A positive Lyapunov exponent indicates exponential divergence, reflecting a strong sensitivity to initial conditions and signaling chaotic behaviour. In contrast, a negative exponent implies convergence and denotes stability within the system. Several studies have been conducted to investigate phase transitions in both quantum and classical regimes by analyzing their underlying chaotic dynamics. Significant examples highlighting the interplay of chaos and quantum phase transition include the famous Sachdev-Ye-Kitaev Model (SYK) and its variants \cite{syk,syk2}, Dicke Model \cite{Dicke}, finite Fermi systems and quantum dots \cite{finite}, long range models of coupled oscillators \cite{coscll} etc. In the recent years, Lyapunov exponents has been used to study the chaotic dynamics in the realms of general relativity and black hole physics. Extensive studies on the dynamics of particles in static axis symmetric spacetimes \cite{static}, around the rotating charged Kerr Newmann black holes \cite{kerr,kerr2}, multi black hole configuration \cite{multi}, black holes incorporating quantum gravity corrections \cite{qg,qg2} has been conducted. In their influential work, Maldacena, Shenker and Stanford identified a universal upper bound (MSS bound) on the rate of growth of chaos in quantum systems with a large number of degrees of freedom and a well-defined semiclassical gravity dual \cite{mss}. The bound is expressed in terms of the Lyapunov exponent $\lambda_L$, which characterizes the exponential growth of out-of-time-ordered correlators (OTOCs), a key diagnostic of quantum chaos. The MSS bound, in natural units, is given by the simple expression, $\lambda_L\leq2\pi\tilde{T}$. Subsequent investigations have validated this bound in various contexts such as for the motion of massive particles near black hole horizon \cite{hori}, its intrinsic connection to the presence of horizon \cite{hori2}. However, several studies have reported the bound violation \cite{vio,vio2,vio3,vio4}

The first attempt to probe the thermodynamic phase structure of black holes through Lyapunov exponents was put forward in in ref \cite{first}. The authors conjectured the existence of a relationship between Lyapunov exponents and black hole phase transitions indicating that chaotic dynamics may serve as a diagnostic tool for probing thermodynamic behaviour in black hole spacetimes. They show that the Lyapunov exponent associated with massless and massive particles moving in unstable circular orbits around black holes, show multivaluedness as a function of temperature. The different branches of the Lyapunov exponent corresponds to the different phases of the black hole. Notably, at a specific critical value of a model-dependent parameter, the Lyapunov exponent becomes single-valued. They also calculate the discontinuous change in the Lyapunov exponent and conclude that it can be treated as an order parameter having a critical exponent of $1/2$ near the critical point. Following this, few other works has been carried out that confirmed the results \cite{le,le2,le3,le4,le5,le6,le7,le8}.

In this work, we apply the technique of using Lyapunov exponents to probe the phase transitions of a particular class of black holes known as \textit{R}-charged black holes. We particularly focus on the four-dimensional \textit{R}-charged black hole, which arises as a solution to gauged supergravity theories. These black holes possess multiple $U(1)$ charges associated with \textit{R}-symmetries in the dual gauge theory. The $4D$ \textit{R}-charged black hole possess four charges and the $5D$ and $7D$ possesses three and two charges respectively. The presence of multiple charges leads to a rich and intricate thermodynamic phase structure \cite{r1,r2,r3,r4,r5}, making this black hole an ideal candidate for exploring the dynamical signatures of phase transitions through Lyapunov exponents. Furthermore, its relevance to the AdS/CFT correspondence adds an additional layer of theoretical significance to the investigation \cite{rh}. For instance, \textit{R}-charged black holes have been used as holographic backgrounds to explore the optical properties of strongly coupled media, including negative refractive index and superconducting behaviour, as demonstrated in refs. \cite{pjp,pjp2,Mahapatra}. Additionally, \textit{R}-charged black holes have been used to study the hydrodynamic regime of strongly coupled gauge theories, where quantities such as shear viscosity and diffusion coefficients have been computed holographically using these charged AdS backgrounds \cite{hydror,shear}

The organisation of the paper is as follows. In section \ref{sec1} we provide a brief review of the calculation of Lyapunov exponents. In section \ref{sec2} we review the \textit{R}-charged black holes. In the next section (\ref{sec3})we calculate the Lyapunov exponents for massless and massive particles for different charge configuration of the \textit{R}-charged black holes and analyze the phase transition through their dynamical behaviour. In section \ref{sec4} we calculate the discontinuous change of the Lyapunov exponent and find the critical exponent near the critical point. Finally, in the last section (\ref{sec5}), we summarize the results.

\section{Review of Lyapunov exponents}\label{sec1}
In this section we briefly overview the calculation of the Lyapunov exponents. We consider massive and massless particles in unstable circular orbits on the equatorial plane. To begin with, we write the lagrangian of a particle's geodesic motion with $\theta=\pi/2$
\begin{equation}\label{eq1}
2\mathcal{L}=-f(r)\dot{t}^2+\frac{\dot{r}^2}{f(r)}+r^2\dot{\phi^2}
\end{equation}
where the dot represent a derivative with respect to the proper time $\tau$. The canonical momenta are derived from the Lagrangian using the standard definition $p_q=\frac{\partial\mathcal{L}}{\partial\dot{q}}$, giving us,
\begin{equation}\label{eq2}
p_t=\frac{\partial\mathcal{L}}{\partial\dot{t}}=-f(r)\dot{t}=-E,\; p_r=\frac{\partial\mathcal{L}}{\partial\dot{r}}=\frac{\dot{r}}{f(r)},\; p_\phi=\frac{\partial\mathcal{L}}{\partial\dot{\phi}}=r^2\dot{\phi}=L
\end{equation}
$E$ and $L$ being the conserved energy and angular momentum of the particle. Equation (\ref{eq2}) gives us \begin{equation}\label{eq3}
\dot{t}=\frac{E}{f(r)} \quad \dot{\phi}=\frac{L}{r^2}
\end{equation}
We can write down the hamiltonian as follows
\begin{equation*}
\mathcal{H}=p_t\dot{t}+p_r\dot{r}+p_\phi\dot{\phi}-\mathcal{L}
\end{equation*}
which, using the relations in (\ref{eq3}), simplifies to
\begin{equation}\label{eq4}
2\mathcal{H}=-E\dot{t}+\frac{\dot{r}^2}{f(r)}+\frac{L^2}{r^2}=\delta_1
\end{equation}
where $\delta_1=-1$ for timelike geodesic and $0$ for null geodesic. For the radial motion of a particle, the effective potential is given by,
\begin{equation}\label{eq5}
V_{\text{eff}}=f(r)\left[\frac{L^2}{r^2}+\frac{E^2}{f(r)}-\delta_1\right]
\end{equation}
We can find out the angular momentum in terms of the effective potential by putting $E=0$ in equation (\ref{eq5}) and then plugging it into equation (\ref{eq4}) we express the hamiltonian as 
\begin{equation}\label{eq6}
\mathcal{H}=\frac{V_{\text{eff}}-E^2}{2f(r)}+\frac{f(r)p^2_r}{2}+\frac{\delta_1}{2}
\end{equation}
leading to the equation of motion in proper time configuration as
\begin{equation}\label{eq7}
\dot{r}=\frac{\partial\mathcal{H}}{\partial p_r}=f(r)p_r,\quad \dot{p_r}=\frac{\partial\mathcal{H}}{\partial r}=-\frac{V'_{\text{eff}(r)}}{2f(r)}-\frac{f'(r)p^2_r}{2}+\frac{V_{\text{eff}}-E^2}{2f^2(r)}f'(r)
\end{equation}
with primes denoting the derivative with respect to $r$. We now linearize the equations of motion around the circular orbit $r_0$, and express the resulting linear stability matrix $K$ with respect to the coordinate time $t$ as follows
\begin{equation}\label{eq8}
K = \begin{pmatrix}
0 & \frac{f(r_0)}{\dot{t}} \\
-\frac{V''_{\text{eff}}(r_0)}{2f(r_0)\dot{t}} & 0
\end{pmatrix}
\end{equation}

The eigenvalue of the above matrix gives us the Lyapunov exponent
\begin{equation}\label{eq9}
\lambda=\sqrt{-\frac{V''_{\text{eff}}(r_0)}{2\dot{t}^2}}
\end{equation}
where the $\pm$ sign before the square root has been dropped for simplicity.
The condition for unstable circular geodesic is given by \begin{equation}\label{eq10}
V'_{\text{eff}}(r_0)=0,\quad V''_{\text{eff}}(r_0)<0
\end{equation}
from which we will get the radius of the unstable circular orbit $r_0$.
Using these relations in equation (\ref{eq5}) we find \begin{equation}\label{eq11}
\frac{E}{L}=\frac{\sqrt{f(r_0)}}{r_0}
\end{equation}
and putting this in equation (\ref{eq3}) and using $\delta_1=0$ (for massless particles), we get \begin{equation}\label{eq12}
\dot{t}=\frac{L}{r_0\sqrt{f(r_0)}}
\end{equation}
With this, we can finally get the expression of the Lyapunov exponent (\ref{eq9}) as 
\begin{equation}\label{eq13}
\lambda=\sqrt{-\frac{r^2_0f(r_0)}{2L^2}V''_{\text{eff}}(r_0)}
\end{equation}
Similarly, for massive particles ($\delta_1=1$), we get the relations
\begin{equation}\label{eq14}
E^2=\frac{2f^2(r_0)}{2f(r_0)-r_0f'(r_0)}\quad \text{and} \quad L^2=\frac{r^3_{0}f'(r_0)}{2f(r_0)-r_0f'(r_0)}
\end{equation}
Therefore from equation (\ref{eq3}) we get 
\begin{equation}\label{eq15}
\dot{t}=\frac{1}{\sqrt{f(r_0)-\frac{1}{2}r_0f'(r_0)}}
\end{equation}
And finally we get the formula for the Lyapunov exponent of massive particles from equation (\ref{eq9}) as
\begin{equation}\label{eq16}
\lambda=\frac{1}{2}\sqrt{\left[r_0f'(r_0)-2f(r_0)\right]V''_{\text{eff}}}
\end{equation}

\section{\textit{R}-charged black hole background}\label{sec2}
In this section we give an introduction to the \textit{R}-charged black hole and derive the necessary thermodynamic quantities. As stated earlier, we focus on the four dimensional \textit{R}-charged black hole only. Such black holes in $D=4$, $\mathcal{N}=8$ gauged supergravity has been studied in detail in ref. \cite{Duff}. This class of black hole solutions is characterized by the presence of four independent $U(1)$ \textit{R}-charges, typically denoted as $q_i$, ($i$=1, 2, 3, 4). The black hole metric is expressed as
\begin{equation}\label{eq17}
ds^2=-\left(\prod_{i=1}^{4}H_i\right)^{-\frac{1}{2}}fdt^2+\left(\prod_{i=1}^{4}H_i\right)^{\frac{1}{2}}\left(f^{-1}dr^2+r^2d\Omega_{2,k}\right)
\end{equation}
with \begin{equation}\label{eq18}
f=k-\frac{\mu}{r}+g^2r^2\prod_{i=1}^{4}H_i \quad \text{and} \quad H_i=1+\frac{q_i}{r} \quad (i=1, 2, 3, 4)
\end{equation}
Here we consider the Newton's constant $G_4=\frac{1}{4}$ and the spatial curvature $k=1$ for the present case. With these choices, the ADM mass is of the following form
\begin{equation}\label{eq19}
M=2\mu+\sum_{i=1}^{4}q_i=2\left(r_++r^3_+g^2\prod_{i=1}^{4}H_i\right)+\sum_{i=1}^{4}q_i
\end{equation}
with $r_+$ being the radius of the outer horizon defined as the largest non-negative zero of the $f(r)$ [Eq. \ref{eq18}]. Also $g\equiv\frac{1}{l}$ and it is defined as the inverse of the AdS radius. The entropy is expressed in terms of the horizon radius $r_+$ and $q_i$ as \begin{equation}\label{eq20}
S=4\pi\sqrt{\left(q_1+r_+\right)\left(q_2+r_+\right)\left(q_3+r_+\right)\left(q_4+r_+\right)}
\end{equation}
The Hawking temperature is calculated as 
\begin{equation}\label{eq21}
T=\frac{f'(r_+)}{4\pi}\prod_{i=1}^{4}H^{-\frac{1}{2}}_i(r_+)
\end{equation}
which comes out to be 
\begin{equation}\label{eq22}
T=\frac{r_+^2 \left\{l^2+q_2 q_3+\left(q_2+q_3\right) q_4+q_1 \left(q_2+q_3+q_4\right)\right\}+2 \left(q_1+q_2+q_3+q_4\right) r_+^3-q_1 q_2 q_3 q_4+3 r_+^4}{4 \pi  l^2 r_+ \sqrt{\left(q_1+r_+\right) \left(q_2+r_+\right) \left(q_3+r_+\right) \left(q_4+r_+\right)}}
\end{equation}

Equipped with the relevant thermodynamic quantities, we are now prepared to do a detailed analysis of the thermodynamic behaviour of the four-dimensional \textit{R}-charged black hole, in particular, we use Lyapunov exponents to examine the system's dynamical behaviour and look for possible connections with its thermodynamic phase structure which we present in the subsequent section.

\section{Lyapunov exponents and phase transitions of \textit{R}-charged black holes}\label{sec3}
In this section we study the thermodynamics of four-dimensional \textit{R}-charged black hole and investigate the validity of the conjectured connection between phase transitions and Lyapunov exponents. We consider three distinct charge configurations, following ref. \cite{r5}: (i) all four charges equal, (ii) two equal and two unequal charges, and (iii) all four charges unequal. We believe, studying these different charge configurations gives a broader picture of the black hole’s thermodynamic behaviour and helps us understand how reliably Lyapunov exponents can signal phase transitions under varying conditions.

\subsection{All Equal Charges}
We first consider all four charges to be equal i.e. $q_1=q_2=q_3=q_4=q$. With these consideration, the quantities stated in the previous section in equations (\ref{eq20}) and (\ref{eq22}) simplifies to 
\begin{equation}\label{eq23}
S=4 \pi  \sqrt{(q+r_+)^4} \quad \text{and}\quad T=\frac{r_+^2 \left(l^2+6 q^2+8 q r_++3 r_+^2\right)-q^4}{4 \pi  l^2 r_+ \sqrt{\left(q+r_+\right){}^4}}
\end{equation}
Also, with mass simplifying to 
\begin{equation*}
M=2 \left(\frac{\left(q+r_+\right){}^4}{l^2 r_+}+2 q+r_+\right)
\end{equation*}
we can find the Gibbs free energy from the definition $F=M-TS$, which comes out to be,
\begin{equation}\label{eq24}
F=2 \left(\frac{\left(q+r_+\right){}^4}{l^2 r_+}+r_+\right)-\frac{r_+ \sqrt{\frac{\left(q+r_+\right){}^4}{r_+^4}} \left(l^2 r_+^2-\left(q-3 r_+\right) \left(q+r_+\right){}^3\right)}{l^2 \sqrt{\left(q+r_+\right){}^4}}+4 q
\end{equation}
Using dimensional analysis \cite{first}, we observe that various physical quantities scale as specific powers of $l$, allowing them to be expressed accordingly,
\begin{equation}
\label{eq25}
\tilde{r}_+=r_+/l, \quad \tilde{q}=q/l , \quad \tilde{F}=F/l , \quad  \tilde{T}=T l , \quad \text{and}  \quad \tilde{M}=M/l.
\end{equation}
The location of the critical point can be identified by satisfying the condition for an inflection point,
\begin{equation}\label{eq26}
\frac{\partial\tilde{T}}{\partial \tilde{r}_+}=\frac{\partial^2\tilde{T}}{\partial^2 \tilde{r}_+}=0,
\end{equation}
Solving these two equation simultaneously using the Hawking temperature we get the numerical critical values as
\begin{equation}\label{eq27}
\tilde{r}_{+c}=0.226135, \quad \tilde{q}_{c}=0.106975, \quad \tilde{T}_c=0.229159
\end{equation}

\begin{figure}[h!]
	\centerline{
	\includegraphics[scale=.8]{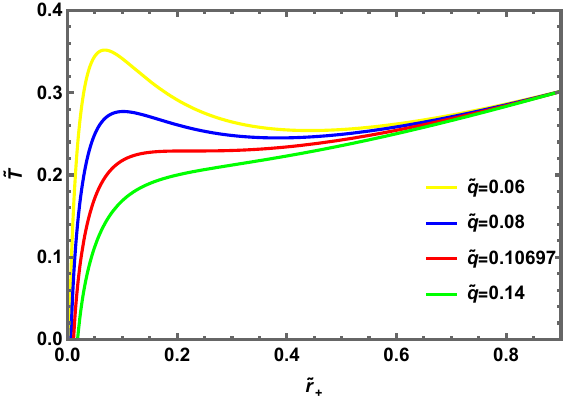}}
	\caption	{Hawking temperature as a function of horizon radius for different values of $\tilde{q}$ with $\tilde{q}_c=0.106975$ (red).}	
	\label{f1}
	\end{figure}
\begin{figure}[h!]
    \centering
    \begin{subfigure}[b]{0.45\textwidth}
        \centering
        \includegraphics[width=\textwidth]{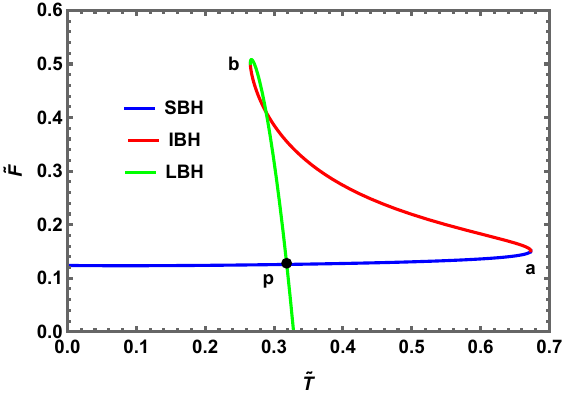}
        \caption{$\tilde{q}=0.09<\tilde{q}_f$}
        \label{f2a}
    \end{subfigure}
    
    \vskip\baselineskip  % Space between rows

    % Bottom two figures
    \begin{subfigure}[b]{0.45\textwidth}
        \centering
        \includegraphics[width=\textwidth]{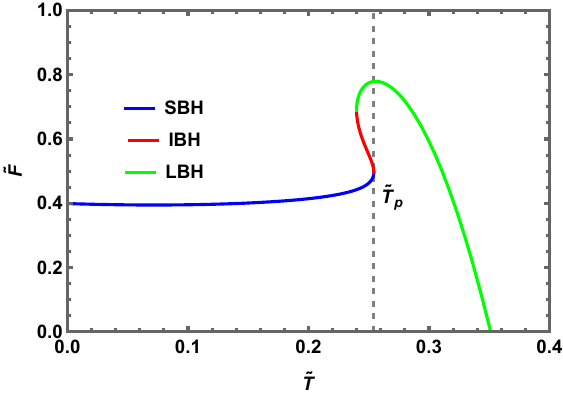}
        \caption{$\tilde{q}_f<\tilde{q}<\tilde{q}_c$}
        \label{f2b}
    \end{subfigure}
    \hfill
    \begin{subfigure}[b]{0.45\textwidth}
        \centering
        \includegraphics[width=\textwidth]{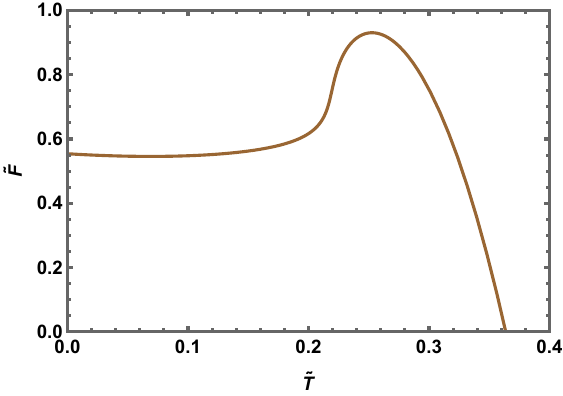}
        \caption{$\tilde{q}>\tilde{q}_c$}
        \label{f2c}
    \end{subfigure}

    \caption{Gibbs free energy as a function of temperature}
    \label{f2}
\end{figure}

We present the plot of Hawking temperature with horizon radius for different values of $\tilde{q}$ in Fig \ref{f1}. Here the yellow and blue curve are for $\tilde{q}<\tilde{q}_c$, the green curve for $\tilde{q}>\tilde{q}_c$ and the red curve corresponds to $\tilde{q}=\tilde{q}_c$. We can observe that for $\tilde{q}$ values less than the critical value, there exists three black hole branches - Small Black Hole (SBH), Intermediate Black Hole (IBH) and Large Black Hole (LBH) branch. We show these branches explicitly in the free energy versus temperature plot in Fig \ref{f2}.
From Fig. \ref{f2a} we observe that it exhibits the well known swallow-tail behaviour which is generally the characteristic of a first-order phase transition. Here the three black hole solution coexists for $\tilde{T}_b<\tilde{T}<\tilde{T}_a$, $\tilde{T}_b$ and $\tilde{T}_a$ being the temperature at the points $b$ and $a$ respectively. At point $p$, with temperature $\tilde{T}_p$, there is a first-order phase transition between the small and the large black hole. Moreover, the free energy associated with the intermediate black hole phase consistently remains higher than that of the small and large black hole phases. This renders the IBH phase thermodynamically disfavored, indicating its instability in the phase structure. Interestingly, when we keep on increasing the value of $\tilde{q}$ towards $\tilde{q}_c$, at a particular of $\tilde{q}=\tilde{q}_f<\tilde{q}_c$, the swallow-tail disappears and there is a finite jump in the free energy at $\tilde{T}=\tilde{T}_p$. This discontinuity in Gibbs free energy demonstrates a zeroth-order phase transition. Such a zeroth-order phase transition has previously been explored in the context of superfluidity and superconductivity \cite{0th}. In the context of black hole thermodynamics, zeroth-order phase transitions have also been reported in several cases \cite{0th2,0th2,0th3,0th4,0th5,0th6,0th7}. As the value of $\tilde{q}$ is increased beyond the critical value $\tilde{q}_c$, the discontinuity vanishes, and the free energy curve becomes continuous. At the critical value, the line of zeroth-order phase transition terminates and the nature of transition changes - it becomes second-order. Now, we investigate the Lyapunov exponents associated with both massless and massive particles moving in unstable circular orbit.
\subsubsection{\textbf{Massless particles}}
\begin{figure}[htbp]
    \centering
    \begin{subfigure}[b]{0.45\textwidth}
        \includegraphics[width=\textwidth]{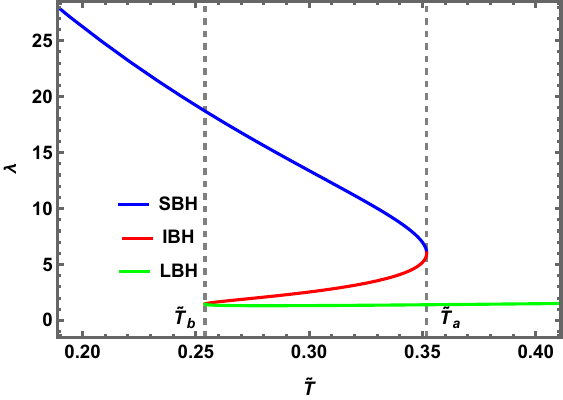}
        \caption{$\tilde{q}=0.06<\tilde{q}_c$}
        \label{f3a}
    \end{subfigure}
    \hspace{0.05\textwidth}
    \begin{subfigure}[b]{0.45\textwidth}
        \includegraphics[width=\textwidth]{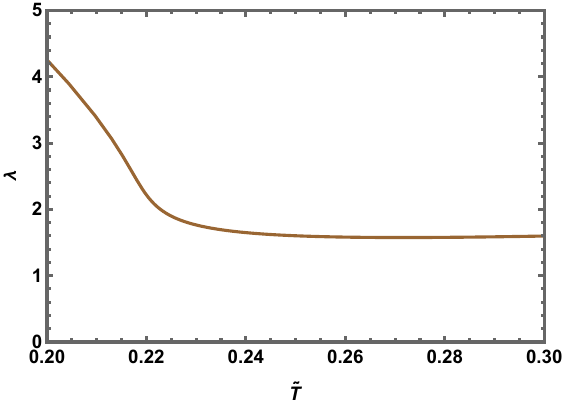}
        \caption{$\tilde{q}>\tilde{q}_c$}
        \label{f3b}
    \end{subfigure}
    
    \vspace{0.5cm} % space between top and bottom rows

    \begin{subfigure}[b]{0.45\textwidth}
        \includegraphics[width=\textwidth]{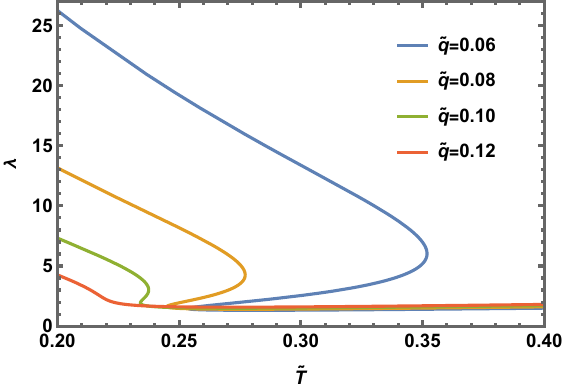}
        \caption{Lyapunov exponent versus temperature}
        \label{f3c}
    \end{subfigure}
    \hspace{0.05\textwidth}
    \begin{subfigure}[b]{0.45\textwidth}
        \includegraphics[width=\textwidth]{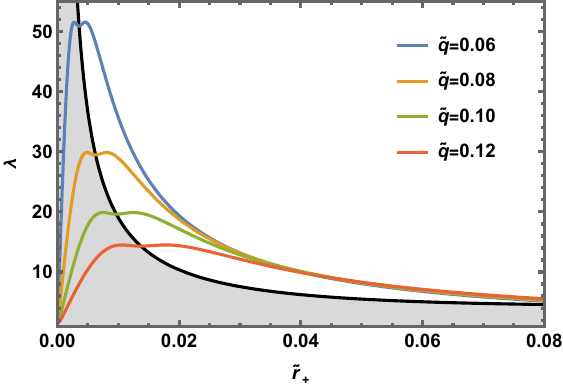}
        \caption{Lyapunov exponent versus horizon radius}
        \label{f3d}
    \end{subfigure}
    \caption{Lyapunov exponent $\lambda$ of massless particles as a function of temperature and horizon radius}
    \label{f3}
\end{figure}

We first calculate the Lyapunov exponent for massless particles. This is done with the help of equations (\ref{eq5}) and (\ref{eq13}). We provide the plot of Lyapunov exponent as a function of temperature of the black hole in Fig. \ref{f3a} where we use the same colour coding as used in Fig. \ref{f2}

For $\tilde{q}$ less than the critical value $\tilde{q}_c$, the Lyapunov exponent shows multivaluedness in the temperature range $\tilde{T}_b$ to $\tilde{T}_a$. As the temperature is increased, the Lyapunov exponent keeps on decreasing. When the temperature reaches $\tilde{T}_b$, $\lambda$ becomes multivalued and the multivaluedness persists till the temperature $\tilde{T}_a$. In the temperature range $\tilde{T}_b<\tilde{T}<\tilde{T}_a$, we observe three distinct branches of the Lyapunov exponent: the blue and green branch corresponds to the small and large branch of the black hole and the red one corresponds to the intermediate black hole branch. This behaviour of Lyapunov exponent is similar to how the Gibbs free energy changes, suggesting it could be useful for identifying black hole phase transitions. We have to mention here that when the transition is first-order (when $\tilde{q}<\tilde{q}_f$), the corresponding temperature $\tilde{T}_p$ lies in between $\tilde{T}_b$ and $\tilde{T}_a$. But for the zeroth order phase transition (when $\tilde{q}_f<\tilde{q}<\tilde{q}_c$), the corresponding temperature coincides exactly with $\tilde{T}_a$. When $\tilde{q}>\tilde{q}_c$, there is no phase transition (Fig. \ref{f2c}) and correspondingly, the Lyapunov exponent also becomes single-valued for $\tilde{q}>\tilde{q}_c$ for all values of temperature as depicted in Fig. \ref{f3b}.

The thermal profile of the Lyapunov exponent for different values of $\tilde{q}$, both above and below $\tilde{q}_c$ are shown in Fig. \ref{f3c}. We can see that the Lyapunov exponent becomes multivalued only for values of $\tilde{q}$ that are less that the critical value. Notably, the way the Lyapunov exponent changes—shifting from multiple values to a single value—accurately signals the location of the second-order critical point $\tilde{q}_c$. To gain better insights on the behaviour of the Lyapunov exponent, we have shown the plot of $\lambda$ with the horizon radius $r_+$. Here the grey area represents non physical region because in this region the Hawking temperature is negative and on the black line it is exactly zero ($\tilde{T}=0$). We can observe that the maximum of $\lambda$ increases as the values of $\tilde{q}$ is decreased. Furthermore, for all values of $\tilde{q}$, the Lyapunov exponent converges to the same value at large $r_+$.

\subsubsection{\textbf{Massive particles}}
\begin{figure}[htbp]
    \centering
    \begin{subfigure}[b]{0.45\textwidth}
        \includegraphics[width=\textwidth]{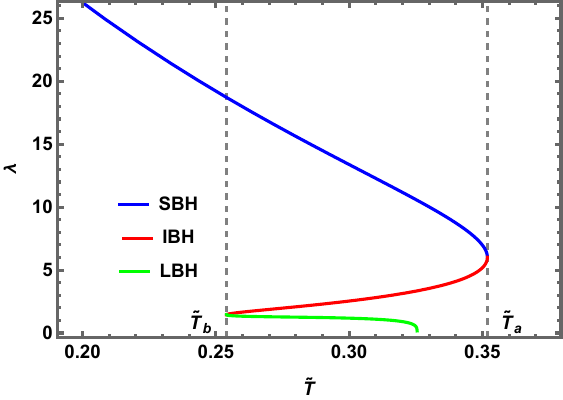}
        \caption{$\tilde{q}=0.06<\tilde{q}_c$}
        \label{f4a}
    \end{subfigure}
    \hspace{0.05\textwidth}
    \begin{subfigure}[b]{0.45\textwidth}
        \includegraphics[width=\textwidth]{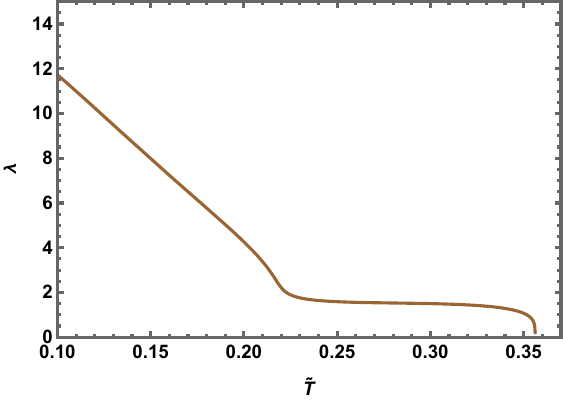}
        \caption{$\tilde{q}>\tilde{q}_c$}
        \label{f4b}
    \end{subfigure}
    
    \vspace{0.5cm} % space between top and bottom rows

    \begin{subfigure}[b]{0.45\textwidth}
        \includegraphics[width=\textwidth]{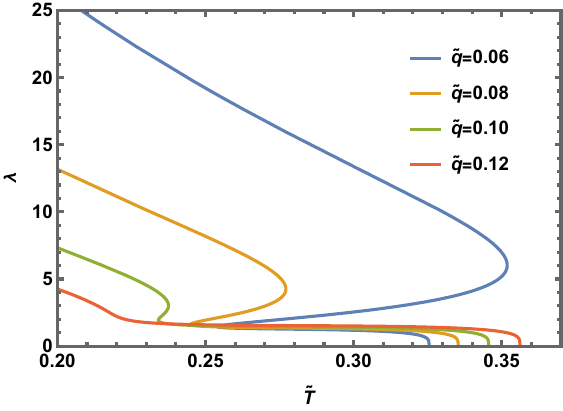}
        \caption{Lyapunov exponent versus temperature}
        \label{f4c}
    \end{subfigure}
    \hspace{0.05\textwidth}
    \begin{subfigure}[b]{0.45\textwidth}
        \includegraphics[width=\textwidth]{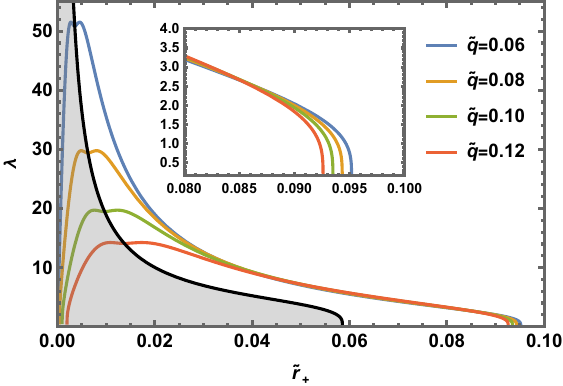}
        \caption{Lyapunov exponent versus horizon radius}
        \label{f4d}
    \end{subfigure}
    \caption{Lyapunov exponent $\lambda$ of massive particles as a function of temperature and horizon radius}
    \label{f4}
\end{figure}

We now compute the Lyapunov exponent associated with massive particles. This is done with the aid of equation (\ref{eq5}) and (\ref{eq16}). We show the variation of Lyapunov exponent of massive particle in Fig. \ref{f4}
For the massive particle case, $r_0$ and $\lambda$ depends on the angular momentum $L$. The plots have been generated using $L=20$ as a representative value for illustration. In the case of massive particles as well, we observe that the Lyapunov exponent displays multivalued behavior when the charge parameter satisfies $\tilde{q}<\tilde{q}_c$ in the temperature range $\tilde{T}_b<\tilde{T}<\tilde{T}_a$ (Fig. \ref{f4a}) and as the charge parameter is increased beyond the critical value, $\tilde{q}>\tilde{q}_c$ this multivaluedness disappears entirely, and the Lyapunov exponent becomes a smooth, single-valued function of temperature (Fig. \ref{f4b}). However, one contrasting feature in the case of massive particle is that here the Lyapunov exponent approaches zero in the large black hole phase.  This is true for all values of $\tilde{q}$ as is illustrated in Fig. \ref{f4c}. In Fig. \ref{f4d} we show the variation of Lyapunov exponent for massive particles with the horizon radius. Here also, the qualitative behaviour mirrors that of massless particles case: maximum of $\lambda$ increases as the values of $\tilde{q}$ decreases.Additionaly, the Lyapunov exponent approaches zero for some value of the horizon radius. Also, the horizon radius where $\lambda$ tends to zero is almost the same for all values of $\tilde{q}$. This suggests that the motion of the massive particles becomes nonchaotic in the large black hole phase. $\lambda$ approaching zero reflects the disappearance of unstable equilibrium points at those horizon radii. This is in contrast to the massless particle case where the Lyapunov exponent attains a non zero value for larger horizon radius.

We thus conclude this section by affirming that the thermodynamic phase transition of the \textit{R}-charged black hole with all four charges equal has been effectively probed using the Lyapunov exponents corresponding to both massless and massive particles in unstable circular orbits.

\subsection{Two Equal and Two Unequal Charges}
\begin{figure}[h!]
	\centerline{
	\includegraphics[scale=.8]{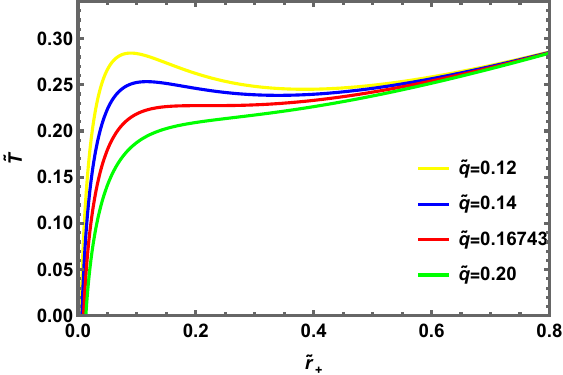}}
	\caption	{Hawking temperature as a function of horizon radius for different values of $\tilde{q}$ with $\tilde{q}_c=0.167433$ (red).}	
	\label{f5}
	\end{figure}

\begin{figure}[h!]
    \centering
    \begin{subfigure}[b]{0.45\textwidth}
        \centering
        \includegraphics[width=\textwidth]{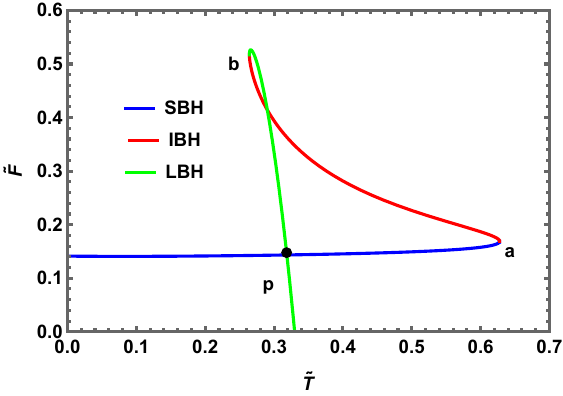}
        \caption{$\tilde{q}=0.05<\tilde{q}_f$}
        \label{f6a}
    \end{subfigure}
    
    \vskip\baselineskip  % Space between rows

    % Bottom two figures
    \begin{subfigure}[b]{0.45\textwidth}
        \centering
        \includegraphics[width=\textwidth]{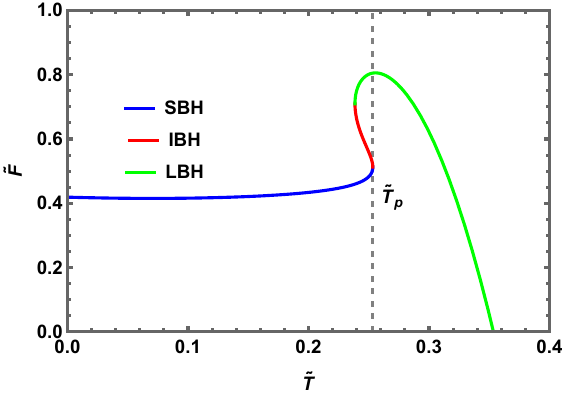}
        \caption{$\tilde{q}_f<\tilde{q}<\tilde{q}_c$}
        \label{f6b}
    \end{subfigure}
    \hfill
    \begin{subfigure}[b]{0.45\textwidth}
        \centering
        \includegraphics[width=\textwidth]{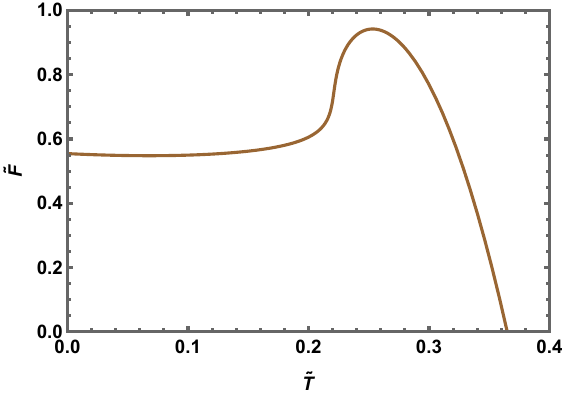}
        \caption{$\tilde{q}>\tilde{q}_c$}
        \label{f6c}
    \end{subfigure}

    \caption{Gibbs free energy as a function of temperature}
    \label{f6}
\end{figure}

We now extend our analysis to the case where two of the four charges of the \textit{R}-charged black hole are equal, while the remaining two are unequal. Specifically, we consider the charge configuration: $q_1=q_2=q$, $q_3=\frac{q}{2}$ and $q_4=\frac{q}{4}$.
Hence, the entropy and temperature given in equations (\ref{eq20}) and (\ref{eq22}) becomes, \begin{equation}\label{eq28}
S=\sqrt{2} \pi  \sqrt{\left(q+r_+\right){}^2 \left(q+2 r_+\right) \left(q+4 r_+\right)}\quad
\end{equation}
\begin{equation}\label{eq29}
T=\frac{8 l^2 r_+^2-\left(q+r_+\right) \left(q+3 r_+\right) \left(q^2-4 q r_+-8 r_+^2\right)}{8 \sqrt{2} \pi  l^2 r_+ \sqrt{\left(q+r_+\right){}^2 \left(q+2 r_+\right) \left(q+4 r_+\right)}}
\end{equation}
The expression of mass simplifies to 
\begin{equation}\label{eq30}
M=\frac{1}{4} \left(\frac{\left(q+2 r_+\right) \left(q+4 r_+\right) \left(q+r_+\right){}^2}{l^2 r_+}+11 q+8 r_+\right)
\end{equation}
With these quantities, the Gibbs free energy can be easily calculated
\begin{multline}\label{eq31}
F=\frac{\sqrt{\left(q+r_+\right){}^2 \left(q+2 r_+\right) \left(q+4 r_+\right)} \left(q^4-r_+^2 \left(8 l^2+21 q^2+44 q r_++24 r_+^2\right)\right)}{8 l^2 r_+^3 \sqrt{\frac{\left(q+r_+\right){}^2 \left(q+2 r_+\right) \left(q+4 r_+\right)}{r_+^4}}}\\+\frac{1}{4} \left(\frac{\left(q+2 r_+\right) \left(q+4 r_+\right) \left(q+r_+\right){}^2}{l^2 r_+}+11 q+8 r_+\right)
\end{multline}
Using the scaling relations provided in equation (\ref{eq25}) along with equation (\ref{eq26}), we compute the numerical values of the critical points.
\begin{equation}\label{eq32}
\tilde{r}_{+c}=0.215966, \quad \tilde{q}_{c}=0.167433, \quad \tilde{T}_c=0.227425
\end{equation}
The variation of the Hawking temperature with horizon radius for different values of $\tilde{q}$ is shown in Fig. \ref{f5}. The yellow and blue curves correspond to values of $\tilde{q}$ less than the critical value $\tilde{q}_c$. In these cases, the temperature profile exhibits two turning points, indicating the presence of three distinct black hole branches. The red curve represents the critical case $\tilde{q} = \tilde{q}_c$, while the green curve, corresponding to $\tilde{q} > \tilde{q}_c$, displays a single smooth branch without turning points.
The plot of the Gibbs free energy as a function of temperature is presented in Fig. \ref{f6}. The three distinct branches of the black hole--small, intermediate, and large--are clearly visible. The characteristic swallow-tail structure indicates the occurrence of a first-order phase transition between the small and large black hole branches. This transition takes place at the point $p$, where the transition temperature is denoted by $\tilde{T}_p$.
Interestingly, for this particular charge configuration too, we observe a zeroth-order phase transition for some value of $\tilde{q}_f<\tilde{q}<\tilde{q}_c$ characterised by the finite jump in the free energy as temperature $\tilde{T}_p$ as depicted in Fig. \ref{f6b}. At the critical value ($\tilde{q}=\tilde{q}_c$), the line of zeroth-order phase transition terminates and the nature of transition changes to second-order. In Fig. \ref{f6c}, the free energy curve for $\tilde{q} > \tilde{q}_c$ is displayed, clearly illustrating that the curve is continuous and single-valued, indicating the absence of any phase transition in this regime.

\subsubsection{\textbf{Massless particles}}
\begin{figure}[htbp]
    \centering
    \begin{subfigure}[b]{0.45\textwidth}
        \includegraphics[width=\textwidth]{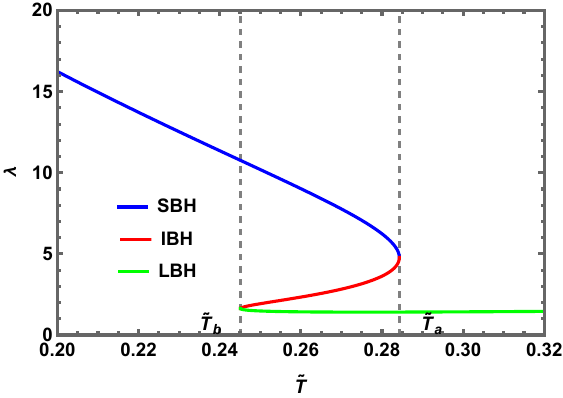}
        \caption{$\tilde{q}=0.12<\tilde{q}_c$}
        \label{f7a}
    \end{subfigure}
    \hspace{0.05\textwidth}
    \begin{subfigure}[b]{0.45\textwidth}
        \includegraphics[width=\textwidth]{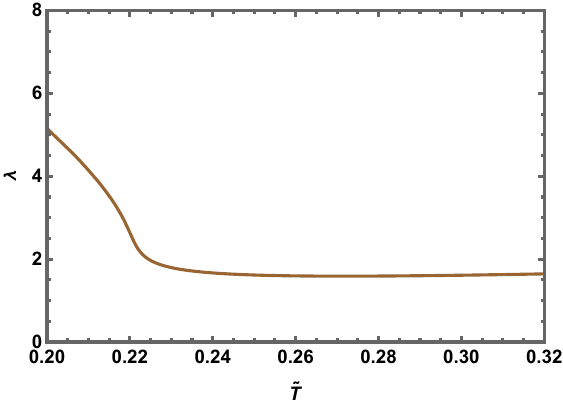}
        \caption{$\tilde{q}>\tilde{q}_c$}
        \label{f7b}
    \end{subfigure}
    
    \vspace{0.5cm} % space between top and bottom rows

    \begin{subfigure}[b]{0.45\textwidth}
        \includegraphics[width=\textwidth]{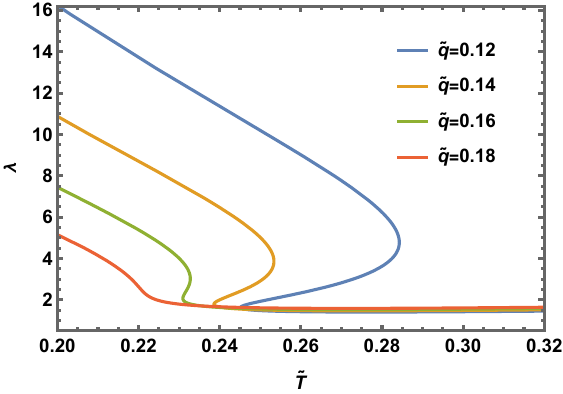}
        \caption{Lyapunov exponent versus temperature}
        \label{f7c}
    \end{subfigure}
    \hspace{0.05\textwidth}
    \begin{subfigure}[b]{0.45\textwidth}
        \includegraphics[width=\textwidth]{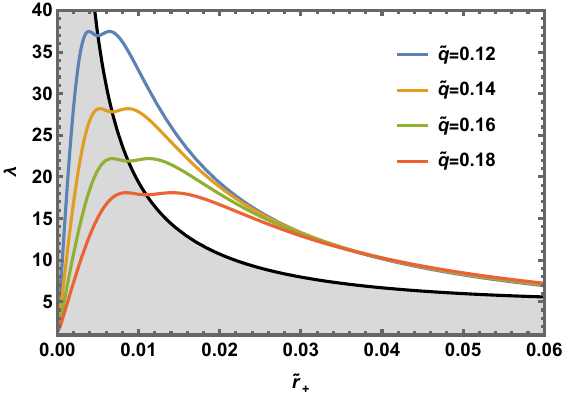}
        \caption{Lyapunov exponent versus horizon radius}
        \label{f7d}
    \end{subfigure}
    \caption{Lyapunov exponent $\lambda$ of massless particles as a function of temperature and horizon radius}
    \label{f7}
\end{figure}

With the charge configuration considered in the previous section, we now proceed to calculate the Lyapunov exponent for massless particles. The variation of the Lyapunov exponent with the Hawking temperature and horizon radius is shown in Fig. \ref{f7}.
Fig. \ref{f7a} show the profile of the Lyapunov exponent with temperature for $\tilde{q}<\tilde{q}_c$ where we observe the multivalued nature of $\lambda$ within the temperature range $\tilde{T}_b<\tilde{T}<\tilde{T}_a$. The three distinct branches of the Lyapunov exponent correspond to the small, intermediate, and large black hole phases, indicated by consistent colour coding. For $\tilde{q}<\tilde{q}_f$, the phase transition temperature lies between $\tilde{T}_a$ and $\tilde{T}_b$ but for $\tilde{q}_f<\tilde{q}<\tilde{q}_c$, $\tilde{T}_p$ coincides with $\tilde{T}_a$. Fig. \ref{f7c} illustrates the variation of $\lambda$ with $\tilde{T}$ for different values of $\tilde{q}$, while Fig. \ref{f7b} presents the behaviour for $\tilde{q}>\tilde{q}_c$. These plots clearly demonstrate that the multivalued nature of the Lyapunov exponent occurs only for $\tilde{q}$ below the critical value, thereby reinforcing its potential role as a diagnostic tool for identifying phase transitions. In Fig. \ref{f7d} we show the variation of Lyapunov exponent with the horizon radius. The grey area represents the non physical region with the Hawking temperature being negative there. In this case as well we observe, like the previously taken configuration, that for all values of $\tilde{q}$, the Lyapunov exponents converge to the same value for large $r_+$.

\subsubsection{\textbf{Massive particles}}
\begin{figure}[htbp]
    \centering
    \begin{subfigure}[b]{0.45\textwidth}
        \includegraphics[width=\textwidth]{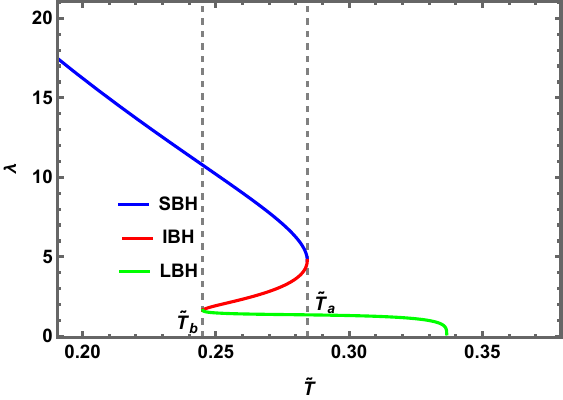}
        \caption{$\tilde{q}=0.12<\tilde{q}_c$}
        \label{f8a}
    \end{subfigure}
    \hspace{0.05\textwidth}
    \begin{subfigure}[b]{0.45\textwidth}
        \includegraphics[width=\textwidth]{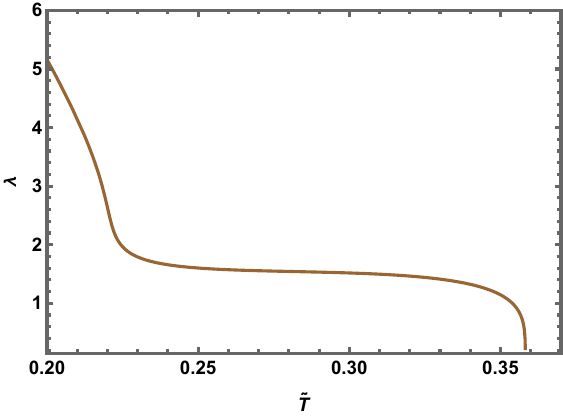}
        \caption{$\tilde{q}>\tilde{q}_c$}
        \label{f8b}
    \end{subfigure}
    
    \vspace{0.5cm} % space between top and bottom rows

    \begin{subfigure}[b]{0.45\textwidth}
        \includegraphics[width=\textwidth]{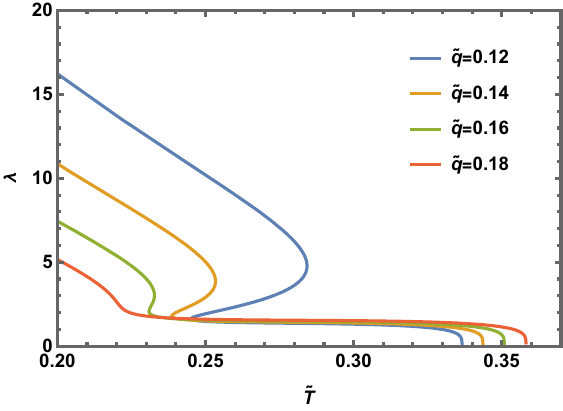}
        \caption{Lyapunov exponent versus temperature}
        \label{f8c}
    \end{subfigure}
    \hspace{0.05\textwidth}
    \begin{subfigure}[b]{0.45\textwidth}
        \includegraphics[width=\textwidth]{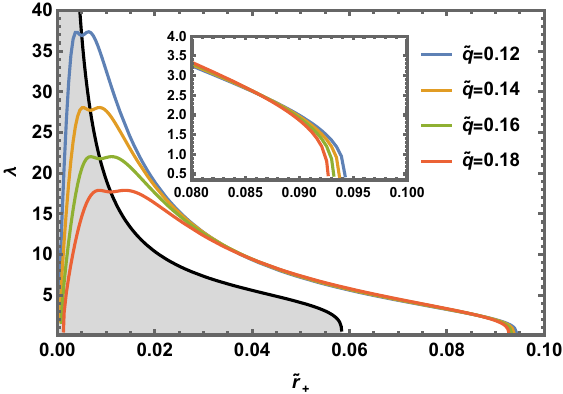}
        \caption{Lyapunov exponent versus horizon radius}
        \label{f8d}
    \end{subfigure}
    \caption{Lyapunov exponent $\lambda$ of massive particles as a function of temperature and horizon radius}
    \label{f8}
\end{figure}

We again compute the Lyapunov exponent associated with massive particle for the same charge configuration (two equal, two unequal). Without bothering about the cumbersome equations, we directly present the thermal profile of the Lyapunov exponent in Fig. \ref{f8a}.
The multivalued nature of the Lyapunov exponent is also observed in the case of massive particles, with its three branches corresponding to the small, intermediate, and large black hole phases. As with the massless case, this multivalued behaviour occurs only for $\tilde{q}$ values less than the critical value $\tilde{q}_c$, as clearly illustrated in Fig. \ref{f8b} and Fig. \ref{f8c}. Moreover, the Lyapunov exponent approaches zero in the large black hole phase at a certain temperature for all values of $\tilde{q}$. This behaviour is also evident from Fig. \ref{f8d}, where $\lambda$ approaches zero at a specific horizon radius. Notably, this value of $r_+$ at which $\lambda$ vanishes remains nearly the same across different $\tilde{q}$ values. These results show that Lyapunov exponents can reliably reflect the phase behaviour of the black hole, even when the charge setup is changed.

\subsection{All Unequal Charges}
\begin{figure}[h!]
	\centerline{
	\includegraphics[scale=.8]{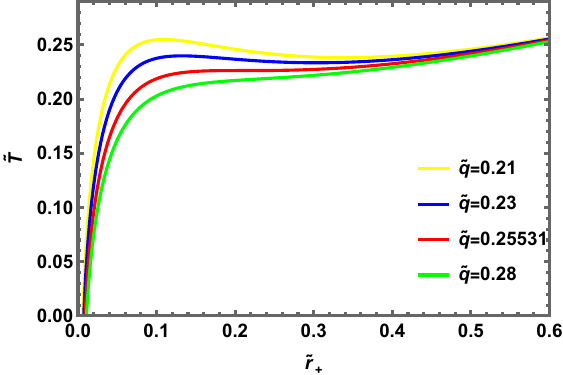}}
	\caption	{Hawking temperature as a function of horizon radius for different values of $\tilde{q}$ with $\tilde{q}_c=0.255316$ (red).}	
	\label{f9}
	\end{figure}

\begin{figure}[h!]
    \centering
    \begin{subfigure}[b]{0.45\textwidth}
        \centering
        \includegraphics[width=\textwidth]{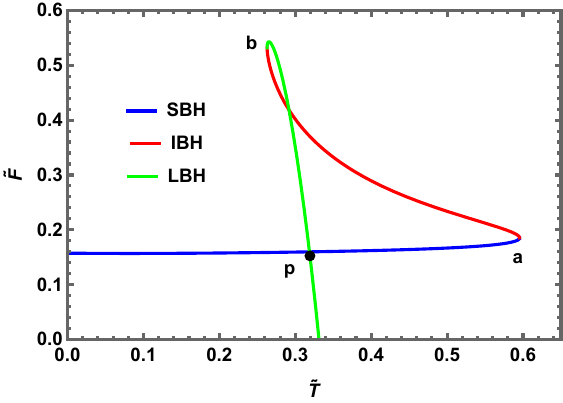}
        \caption{$\tilde{q}=0.05<\tilde{q}_f$}
        \label{f10a}
    \end{subfigure}
    
    \vskip\baselineskip  % Space between rows

    % Bottom two figures
    \begin{subfigure}[b]{0.45\textwidth}
        \centering
        \includegraphics[width=\textwidth]{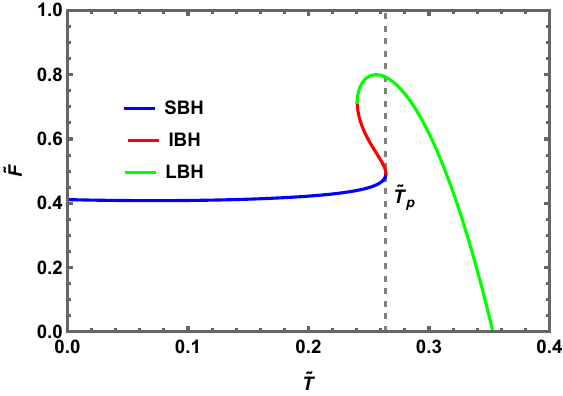}
        \caption{$\tilde{q}_f<\tilde{q}<\tilde{q}_c$}
        \label{f10b}
    \end{subfigure}
    \hfill
    \begin{subfigure}[b]{0.45\textwidth}
        \centering
        \includegraphics[width=\textwidth]{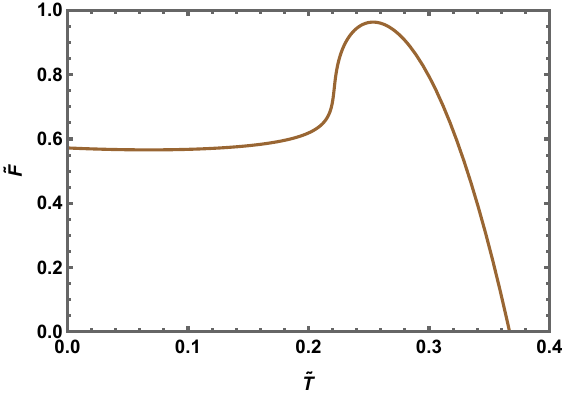}
        \caption{$\tilde{q}>\tilde{q}_c$}
        \label{f10c}
    \end{subfigure}

    \caption{Gibbs free energy as a function of temperature}
    \label{f10}
\end{figure}

We finally take the \textit{R}-charged black hole configuration with all the four charges being unequal. Specifically, we consider $q_1=,q, q_2=\frac{q}{2}, q_3=\frac{q}{4}, q_4=\frac{q}{6}$. In this charge configuration the entropy and the temperature becomes

\begin{equation}\label{eq33}
S=\frac{\pi  \sqrt{\left(q+r_+\right) \left(q+2 r_+\right) \left(q+4 r_+\right) \left(q+6 r_+\right)}}{\sqrt{3}}
\end{equation}
\begin{equation}\label{eq34}
T=\frac{\sqrt{3} \left(r_+^2 \left(l^2+\frac{7 q^2}{6}\right)-\frac{q^4}{48}+\frac{23 q r_+^3}{6}+3 r_+^4\right)}{\pi  l^2 r_+ \sqrt{\left(q+r_+\right) \left(q+2 r_+\right) \left(q+4 r_+\right) \left(q+6 r_+\right)}}
\end{equation}
The mass expression simplifies to
\begin{equation}\label{eq35}
M=\frac{r_+ \left(8 r_+ \left(6 l^2+7 q^2\right)+46 l^2 q+13 q^3+92 q r_+^2+48 r_+^3\right)+q^4}{24 l^2 r_+}
\end{equation}
The free energy is calculated using $F=M-TS$. The expression is too large and not quite intuitive so we just present its variation with the Hawking temperature. But first we analyse the behaviour of Hawking temperature with horizon radius as shown in Fig. \ref{f9}. Here, again we have used the scaling given in equation (\ref{eq25}) and calculated the critical points with the relation (\ref{eq26}) which comes out to be 
\begin{equation}\label{eq36}
\tilde{r}_{+c}=0.209593, \quad \tilde{q}_{c}=0.255316, \quad \tilde{T}_c=0.226348
\end{equation}
Here the yellow and the blue curve corresponds to $\tilde{q}<\tilde{q}_c$. The red and the green curve corresponds to $\tilde{q}=\tilde{q}_c$ and $\tilde{q}>\tilde{q}_c$ respectively. It is observed that for $\tilde{q}$ values less than the critical value, there exists three black hole branches which we explicitly show in Fig. \ref{f10}
For the configuration with all unequal charges as well, we observe the characteristic swallow-tail structure in the free energy, indicating the presence of a first-order phase transition. At point $p$, a first-order phase transition between the small and the large black hole occurs with the transition temperature being $\tilde{T}_p$. Moreover, the free energy associated with the intermediate black hole (IBH) phase consistently remains higher than that of the small and large black hole phases. This renders the IBH phase thermodynamically disfavored, indicating its instability within the overall phase structure. Additionally, a zeroth-order phase transition is also present in this case, marked by a finite jump in the free energy, as illustrated in Fig. \ref{f10b}. At the critical value ($\tilde{q} = \tilde{q}_c$), the line of zeroth-order phase transition terminates, and the nature of the transition changes to second-order. When $\tilde{q}$ becomes greater than the critical value, the free energy becomes a continuous curve (Fig. \ref{f10c}) confirming that no phase transition occurs in this regime.

\subsubsection{\textbf{Massless particles}}
\begin{figure}[htbp]
    \centering
    \begin{subfigure}[b]{0.45\textwidth}
        \includegraphics[width=\textwidth]{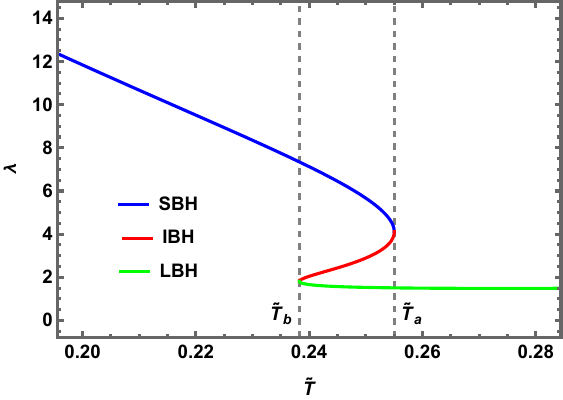}
        \caption{$\tilde{q}=0.21<\tilde{q}_c$}
        \label{f11a}
    \end{subfigure}
    \hspace{0.05\textwidth}
    \begin{subfigure}[b]{0.45\textwidth}
        \includegraphics[width=\textwidth]{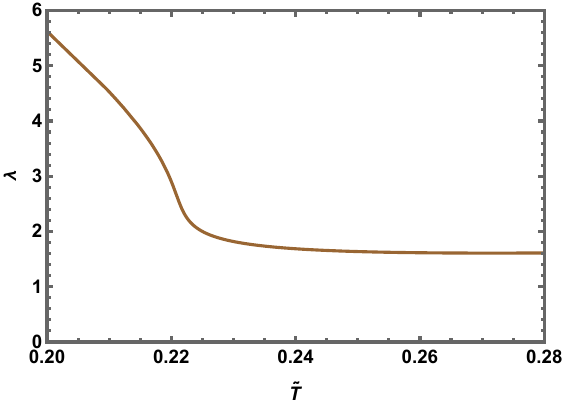}
        \caption{$\tilde{q}>\tilde{q}_c$}
        \label{f11b}
    \end{subfigure}
    
    \vspace{0.5cm} % space between top and bottom rows

    \begin{subfigure}[b]{0.45\textwidth}
        \includegraphics[width=\textwidth]{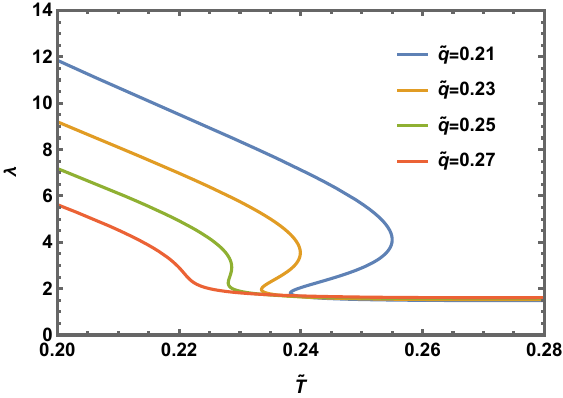}
        \caption{Lyapunov exponent versus temperature}
        \label{f11c}
    \end{subfigure}
    \hspace{0.05\textwidth}
    \begin{subfigure}[b]{0.45\textwidth}
        \includegraphics[width=\textwidth]{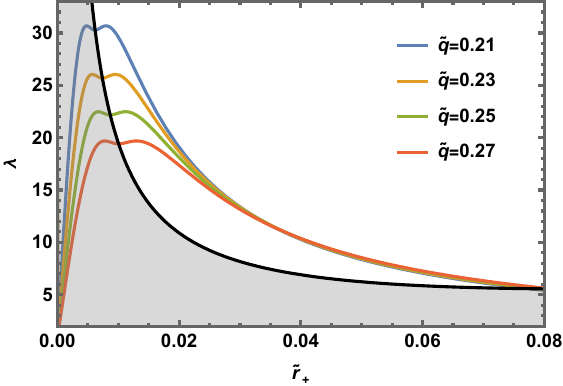}
        \caption{Lyapunov exponent versus horizon radius}
        \label{f11d}
    \end{subfigure}
    \caption{Lyapunov exponent $\lambda$ of massless particles as a function of temperature and horizon radius}
    \label{f11}
\end{figure}

We now compute the Lyapunov exponent associated with massless particles for the all-unequal charge configuration of the \textit{R}-charged black hole. This is carried out using the expressions provided in equations (\ref{eq5}) and (\ref{eq13}). As the resulting expression is quite lengthy and not particularly illuminating in closed form, we instead focus on illustrating its behaviour through its variation with temperature and horizon radius, as shown in Fig. \ref{f11}.
For this particular charge configuration as well, the Lyapunov exponent exhibits a behaviour consistent with the previous cases. Initially, it decreases with increasing Hawking temperature. Within the specific temperature range $\tilde{T}_b < \tilde{T} < \tilde{T}_a$, the multivalued nature of the exponent emerges—where the blue segment corresponds to the small black hole branch, the red to the intermediate (unstable) black hole branch, and the green to the large black hole branch. For $\tilde{q}$ values greater than the critical value, the multivalued nature of the Lyapunov exponent disappears, and it transitions into a smooth, single-valued curve, as shown in Fig. \ref{f11b}. This behavior aligns with what we previously observed in the analyses of free energy and temperature. The variation of the Lyapunov exponent for different values of $\tilde{q}$ is depicted in Fig. \ref{f11c}. For the all-unequal charge configuration as well, the Lyapunov exponent is observed to converge to the same value at large horizon radius, as illustrated in Fig. \ref{f11d} where the grey area is the non physical region with $\tilde{T}$ being negative.

\subsubsection{\textbf{Massive paritcles}}
\begin{figure}[htbp]
    \centering
    \begin{subfigure}[b]{0.45\textwidth}
        \includegraphics[width=\textwidth]{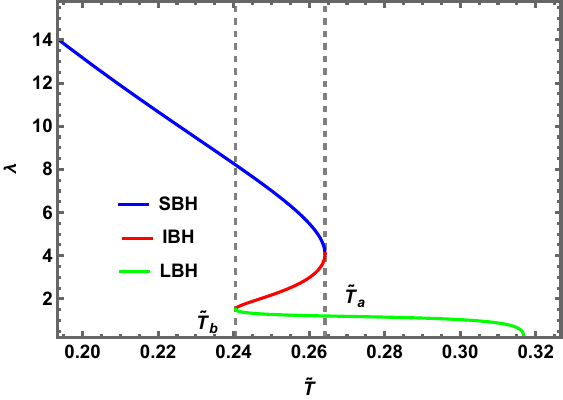}
        \caption{$\tilde{q}=0.20<\tilde{q}_c$}
        \label{f12a}
    \end{subfigure}
    \hspace{0.05\textwidth}
    \begin{subfigure}[b]{0.45\textwidth}
        \includegraphics[width=\textwidth]{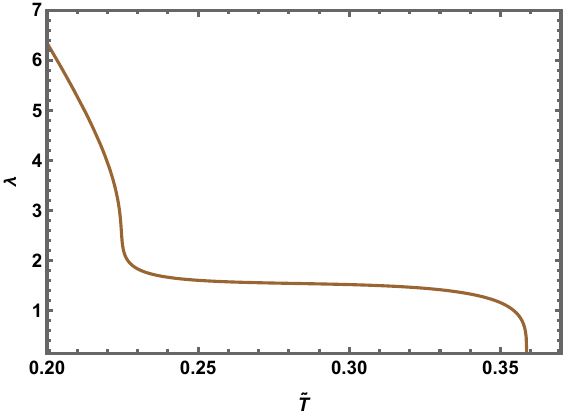}
        \caption{$\tilde{q}>\tilde{q}_c$}
        \label{f12b}
    \end{subfigure}
    
    \vspace{0.5cm} % space between top and bottom rows

    \begin{subfigure}[b]{0.45\textwidth}
        \includegraphics[width=\textwidth]{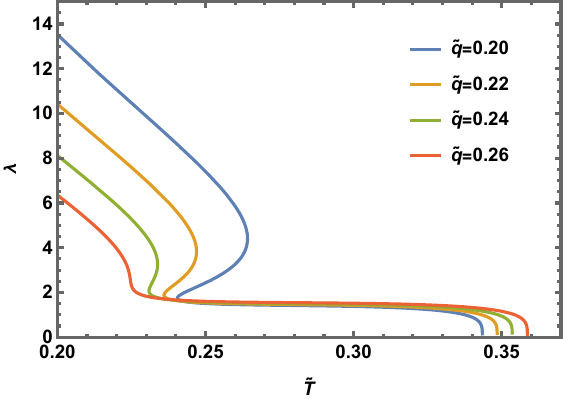}
        \caption{Lyapunov exponent versus temperature}
        \label{f12c}
    \end{subfigure}
    \hspace{0.05\textwidth}
    \begin{subfigure}[b]{0.45\textwidth}
        \includegraphics[width=\textwidth]{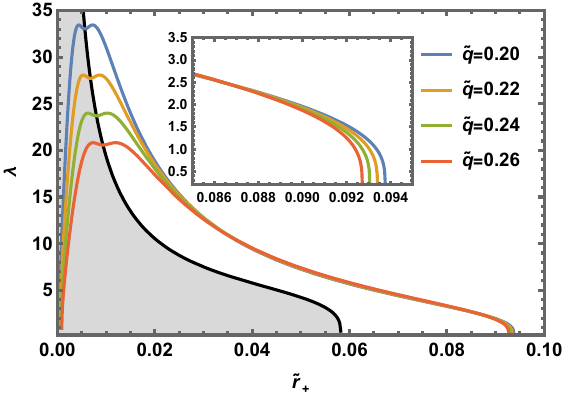}
        \caption{Lyapunov exponent versus horizon radius}
        \label{f12d}
    \end{subfigure}
    \caption{Lyapunov exponent $\lambda$ of massive particles as a function of temperature and horizon radius}
    \label{f12}
\end{figure}

Lastly, we compute the Lyapunov exponent associated with massive particles for the all-unequal charge configuration of the \textit{R}-charged black hole. As with the previously considered charge configurations, this case also allows us to probe the thermodynamic phase transition through the behaviour of the Lyapunov exponent. The thermal profile of the exponent is presented in Fig \ref{f12a}. The multivalued nature of the Lyapunov exponent is clearly observed for $\tilde{q} < \tilde{q}_c$, while it becomes single-valued for $\tilde{q} > \tilde{q}_c$, as is evident from Fig. \ref{f12b} and Fig. \ref{f12c}.
In this charge configuration as well, the Lyapunov exponent for massive particles approaches zero in the large black hole phase for all values of $\tilde{q}$, indicating that the motion becomes nonchaotic in this regime. This vanishing behavior also occurs at specific horizon radii, reflecting the absence of unstable equilibrium points at those locations. 

Thus, for all the three charge configurations considered, we find that the Lyapunov exponent can successfully probe the black hole phase thermodynamic phase transitions. This consistent behaviour highlights the robustness of the Lyapunov exponent as a diagnostic tool. With this, we conclude our analysis of the phase transition of \textit{R}-charged black holes in various charge configurations.

\section{Critical exponents of \textit{R}-charged black holes with lyapunov exponent}\label{sec4}
The concept of order parameter is central in the Landau theory of phase transitions and in the study of critical phenomenon and universality classes in statistical mechanics. It is a measurable quantity that is zero in one phase and non-zero in another, for example, in ferromagnetism the order parameter is the net magnetisation $M$ which is zero for $T>T_c$ and non zero for $T<T_c$. Similarly, the liquid-gas transitions, the order parameter is the differences in densities which is zero at the critical point. The order parameter critical exponent for liquid-gas transition is specifically $1/2$. Remarkably, the transition between the small and large black hole (SBH–LBH) phases can be effectively characterized by the difference in the Lyapunov exponent. At the phase transition temperature $\tilde{T}_p$, we denote the Lyapunov exponent corresponding to the small black hole branch as $\lambda_s$, and that of the large black hole branch as $\lambda_l$. At the second-order critical point, where $\tilde{T}_p = \tilde{T}_c$, and $\lambda_s = \lambda_l = \lambda_c$, with $\lambda_c$ being the critical value of the Lyapunov exponent obtained by evaluating $\lambda$ at the critical values of the relevant thermodynamic quantities. The discontinuity in the Lyapunov exponent $\Delta\lambda=\lambda_s -\lambda_l$ serves as an order parameter with it being zero at the critical point. To study the critical behaviour of $\lambda$, we calculate the critical exponent-a quantity that characterizes how a physical observable behaves near the critical point of a phase transition. For this purpose, we employ the technique outlined in \cite{le2,ce}. We calculate the critical exponent, $\delta$ satisfying 
\begin{equation}\label{eq37}
\Delta\lambda\equiv\lambda_s-\lambda_l=\mid\tilde{T}-\tilde{T}_c\mid^\delta
\end{equation}
First we express the horizon radius and the Hawking temperature at the phase transition point as \begin{equation}\label{eq38}
\tilde{r}_p=\tilde{r}_c\left(1+\Delta\right) \quad \text{and} \quad \tilde{T}(\tilde{r}_+)=\tilde{T}_c\left(1+\epsilon\right)
\end{equation}
where $\mid\Delta\mid\ll1$ and $\mid\epsilon\mid\ll1$. We now Taylor expand the Lyapunov exponent about the critical point of $\tilde{r}_c$
\begin{equation}\label{eq39}
\lambda=\lambda_c+\left[\frac{\partial \lambda}{\partial \tilde{r}_+}\right]d\tilde{r}_++\mathcal{O}\left(\tilde{r}_+\right)
\end{equation}
where we have used the subscript $c$ for representing the values at the critical point. Now, using equation (\ref{eq38}) and (\ref{eq39}) we can find,
\begin{equation}\label{eq40}
\frac{\Delta\lambda}{\lambda_c}=\frac{\lambda_s-\lambda_l}{\lambda_c}=\frac{\tilde{r}_c}{\lambda}_c\left[\frac{\partial\lambda}{\partial\tilde{r}_+}\right]_c\left(\Delta_s-\Delta_l\right)
\end{equation}
where we have also used the fact that at the critical point $\lambda_s = \lambda_l = \lambda_c$ and so $\lambda_s(\tilde{r}_c)-\lambda_l(\tilde{r}_c)=0$.
In a similar way, we also Taylor expand the Hawking temperature about the critical point $\tilde{r}_c$ and obtain 
\begin{equation}\label{eq41}
\tilde{T}=\tilde{T}_c+\frac{\tilde{r}^2_c}{2}\left[\frac{\partial^2 \tilde{T}}{\partial \tilde{r}^2_+}\right]\Delta^2
\end{equation}
where we have omitted the higher order terms and used $\left[\frac{\partial\tilde{T}}{\partial\tilde{r}_+}\right]_c\rightarrow 0$. Finally using the two equations (\ref{eq40}) and (\ref{eq41}), we obtain a simplified expression
\begin{equation}\label{eq42}
\frac{\Delta\lambda}{\lambda_c}=k\sqrt{t-1}
\end{equation}
where $t=\frac{\tilde{T}}{\tilde{T}_c}$ and 
\begin{equation}\label{eq43}
k=\frac{\sqrt{\tilde{T}_c}}{\lambda_c}\left[\frac{\partial\Delta\lambda}{\partial\tilde{r}_+}\right]_c\left[\frac{1}{2}\frac{\partial^2\tilde{T}}{\partial\tilde{r}^2_+}\right]^{-1/2}_c
\end{equation}
Therefore, we conclude that the critical exponent $\delta$ related to the order parameter $\Delta\lambda$ near the critical point in $1/2$

\subsection{Numerical Verification}
\begin{figure}[h!]
    \centering
    \begin{subfigure}[b]{0.45\textwidth}
        \centering
        \includegraphics[width=\textwidth]{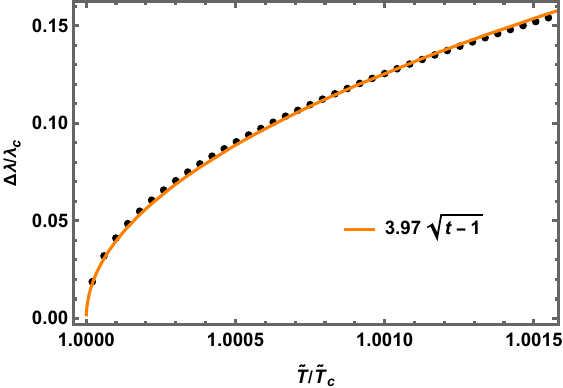}
        \caption{all equal charges}
        \label{f13a}
    \end{subfigure}
    
    \vskip\baselineskip  % Space between rows

    % Bottom two figures
    \begin{subfigure}[b]{0.45\textwidth}
        \centering
        \includegraphics[width=\textwidth]{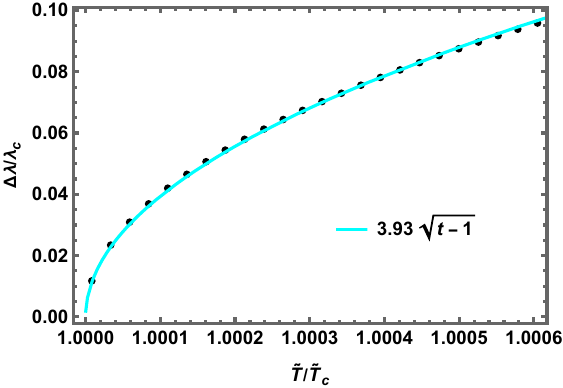}
        \caption{two equal two unequal charges}
        \label{f13b}
    \end{subfigure}
    \hfill
    \begin{subfigure}[b]{0.45\textwidth}
        \centering
        \includegraphics[width=\textwidth]{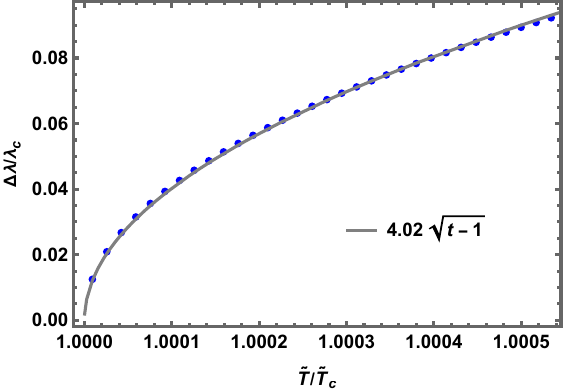}
        \caption{all unequal charges}
        \label{f13c}
    \end{subfigure}

    \caption{Rescaled discontinuity in Lyapunov exponent $\Delta\lambda/\lambda_c$ versus rescaled phase transition temperature $t$ for massless particle case}
    \label{f13}
\end{figure}

We now give a numerical verification of the above mentioned result for the $R$-charged black holes. We do it for both massive and massless particles. We plot the rescaled order parameter $\frac{\Delta\lambda}{\lambda_c}$ versus $\frac{\tilde{T}_p}{\tilde{T}_c}$ denoted by $t$. For massless particles, the plot of $\Delta\lambda/\lambda_c$ vs $t$ for all the three charge configurations are shown in Fig. \ref{f13}. In this figure, the plot is shown only near the critical value. Here the dots are the data points that we have calculated numerically and the solid coloured lines represent the fitted curves.

\begin{figure}[h!]
    \centering
    \begin{subfigure}[b]{0.45\textwidth}
        \centering
        \includegraphics[width=\textwidth]{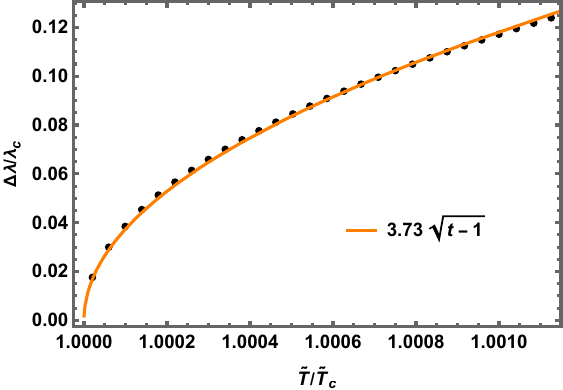}
        \caption{all equal charges}
        \label{f14a}
    \end{subfigure}
    
    \vskip\baselineskip  % Space between rows

    % Bottom two figures
    \begin{subfigure}[b]{0.45\textwidth}
        \centering
        \includegraphics[width=\textwidth]{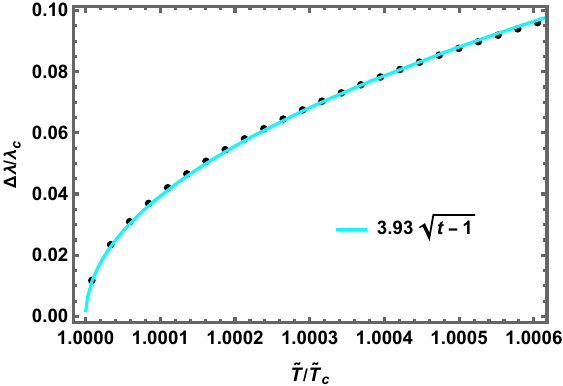}
        \caption{two equal two unequal charges}
        \label{f14b}
    \end{subfigure}
    \hfill
    \begin{subfigure}[b]{0.45\textwidth}
        \centering
        \includegraphics[width=\textwidth]{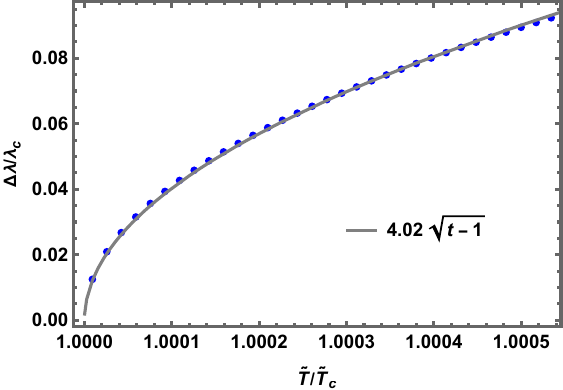}
        \caption{all unequal charges}
        \label{f14c}
    \end{subfigure}

    \caption{Rescaled discontinuity in Lyapunov exponent $\Delta\lambda/\lambda_c$ versus rescaled phase transition temperature $t$ for massive particle case}
    \label{f14}
\end{figure}

We find that the expression $\Delta\lambda/\lambda_c = k\sqrt{t - 1}$ fits our numerically calculated data remarkably well across all considered charge configurations. The value of the constant $k$ varies depending on the specific charge configuration but the critical exponent $\delta$ remains fixed i.e. $1/2$. In Fig. \ref{f14} we provide the plots $\Delta\lambda/\lambda_c$ versus $t$ for different charge configurations. In this case also, the value of the critical exponent $\delta$ remains fixed ($1/2$) while the value of $k$ differs for different charge configurations considered. The corresponding values of $k$ are tabulated in \ref{tbl1}.

\begin{table}[h!]
\centering
\begin{tabular}{|c|c|c|c|}
\hline
\text{Charge Configuration} & \text{Particle Type} & $k$ & critical exponent ($\delta$) \\
\hline
\multirow{2}{*}{All equal} 
    & Massless & 3.97061 & $\frac{1}{2}$ \\
    & Massive  & 3.73783 & $\frac{1}{2}$ \\
\hline
\multirow{2}{*}{Two equal, two unequal} 
    & Massless & 3.93412 & $\frac{1}{2}$ \\
    & Massive  & 3.93947 & $\frac{1}{2}$ \\
\hline
\multirow{2}{*}{All unequal} 
    & Massless & 4.02495 & $\frac{1}{2}$ \\
    & Massive  & 4.02990 & $\frac{1}{2}$ \\
\hline
\end{tabular}
\caption{Values of $k$ and critical exponent $\delta$ for massless and massive particles in various charge configurations}
\label{tbl1}
\end{table}

\section{conclusion and discussion}\label{sec5}
In this work we have extended the investigation of the proposed connection between black hole phase transitions and Lyapunov exponents to the case of \textit{R}-charged black holes in AdS space. We examined three different charge configuration in this paper-all equal charges, two equal and two unequal charges and all unequal charges-to probe the thermodynamic phase transition through the behaviour of Lyapunov exponents associated with unstable circular orbits of both massless and massive particles. We considered the four dimensional \textit{R}-charged black hole for our study not only because of its rich and intricate thermodynamics phase structure, but also due to its immense relevance to the AdS/CFT correspondence-a few of which have been discussed in the section \ref{sec0}.

By analyzing the standard thermodynamics using Gibbs free energy, we found a zeroth-order phase transition characterised by the finite jump in the free energy occurring within the range $\tilde{q}_f<\tilde{q}<\tilde{q}_c$. Below $\tilde{q}_f$ we observe the conventional first-order phase transition characterised by the swallow-tail behaviour. This behaviour is seen across all three charge configurations considered in our study.

We have observed that the thermal profile of the Lyapunov exponent exhibits multivaluedness for $\tilde{q}$ values smaller than the critical value $\tilde{q}_c$. The three branches of the Lyapunov exponent correspond to the three black hole branches-SBH, IBH, LBH. The multivaluedness disappears for $\tilde{q}>\tilde{q}_c$ which is again reminiscent of the free energy behaviour. This behaviour is observed for both massless and massive particles. These observations imply that the Lyapunov exponent carries information about the underlying thermodynamic phase structure of the \textit{R}-charged black hole considered.

We also studied the discontinuous change of the Lyapunov exponent ($\Delta\lambda$) for both massless and massive particle cases and discussed how it can be treated as an order parameter for black hole phase transition. Theoretically, we derived that the critical exponent $\delta$ related to the order parameter is precisely $1/2$ which is the same for van der Waals fluid as predicted by mean-field theory.  This result was further supported numerically, where our analysis of $\Delta\lambda/\lambda_c$ near the critical point showed excellent agreement with the theoretical prediction. It is important to note that, although in our four-dimensional \textit{R}-charged black hole case we observed a zeroth-order phase transition line terminating at a point where the transition becomes second-order, the critical exponent still turns out to be $1/2$. This outcome is expected, as the mean-field universality class is determined by the symmetry and dimensionality of the system near criticality, rather than the nature of the lower-temperature phase transition. Hence, our study further validates the conjectured relationship between Lyapunov exponent and phase transition of black holes, establishing it as a viable probe and effective order parameter. This study also supports the universality of the critical exponent, consistently yielding the value $1/2$.

We have also studied two additional charge configurations--two pairs of equal charges and three equal charges--and found consistent results. However, we do not include them here as the analysis becomes repetitive. We admit that we have not identified a clear signature of the zeroth-order phase transition from the behaviour of the Lyapunov exponent and it would be interesting to explore this aspect further. Additionally, studying black hole phase transitions in different theories of gravity and across various thermodynamic ensembles could provide deeper insight into the robustness of the conjectured connection. We intend to pursue these directions in our future work.


\begin{thebibliography}{99}

\bibitem{Phys}
S.~W.~Hawking,
``Gravitational radiation from colliding black holes,''
Phys. Rev. Lett. \textbf{26}, 1344-1346 (1971)
doi:10.1103/PhysRevLett.26.1344
%1119 citations counted in INSPIRE as of 10 May 2025
%\cite{Bekenstein:2008smd}
\bibitem{bekens}
J.~Bekenstein,
``Bekenstein-Hawking entropy,''
Scholarpedia \textbf{3}, no.10, 7375 (2008)
doi:10.4249/scholarpedia.7375
%19 citations counted in INSPIRE as of 10 May 2025
%\cite{Hawking:1974rv}
\bibitem{Hawking}
S.~W.~Hawking,
``Black hole explosions,''
Nature \textbf{248}, 30-31 (1974)
doi:10.1038/248030a0
%4994 citations counted in INSPIRE as of 10 May 2025
%\cite{Hawking:1975vcx}
\bibitem{Hawking2}
S.~W.~Hawking,
``Particle Creation by Black Holes,''
Commun. Math. Phys. \textbf{43}, 199-220 (1975)
[erratum: Commun. Math. Phys. \textbf{46}, 206 (1976)]
doi:10.1007/BF02345020
%12191 citations counted in INSPIRE as of 10 May 2025
%\cite{Bardeen:1973gs}
\bibitem{Bardeen}
J.~M.~Bardeen, B.~Carter and S.~W.~Hawking,
``The Four laws of black hole mechanics,''
Commun. Math. Phys. \textbf{31}, 161-170 (1973)
doi:10.1007/BF01645742
%3499 citations counted in INSPIRE as of 10 May 2025

%\cite{Davies:1977bgr}
\bibitem{Davies}
P.~C.~W.~Davies,
``Thermodynamics of Black Holes,''
Proc. Roy. Soc. Lond. A \textbf{353}, 499-521 (1977)
doi:10.1098/rspa.1977.0047
%350 citations counted in INSPIRE as of 10 May 2025
\bibitem{Hut}
P. Hut,``Charged black holes and phase transitions,”
Monthly Notices of the Royal
 Astronomical Society, vol. 180, pp. 379–389, 10 1977
 

%\cite{Maldacena:1997re}
\bibitem{Maldacena}
J.~M.~Maldacena,
``The Large N limit of superconformal field theories and supergravity,''
Adv. Theor. Math. Phys. \textbf{2}, 231-252 (1998)
doi:10.4310/ATMP.1998.v2.n2.a1
[arXiv:hep-th/9711200 [hep-th]].
%19437 citations counted in INSPIRE as of 05 Mar 2024

%\cite{Kubiznak:2012wp}
\bibitem{Kubiz}
D.~Kubiznak and R.~B.~Mann,
``P-V criticality of charged AdS black holes,''
JHEP \textbf{07}, 033 (2012)
doi:10.1007/JHEP07(2012)033
[arXiv:1205.0559 [hep-th]].
%1021 citations counted in INSPIRE as of 07 Mar 2024

%\cite{Hawking:1982dh}
\bibitem{Hawkpage}
S.~W.~Hawking and D.~N.~Page,
``Thermodynamics of Black Holes in anti-De Sitter Space,''
Commun. Math. Phys. \textbf{87}, 577 (1983)
doi:10.1007/BF01208266
%2535 citations counted in INSPIRE as of 02 May 2024

%\cite{Cai:2013qga}
\bibitem{Cai}
R.~G.~Cai, L.~M.~Cao, L.~Li and R.~Q.~Yang,
``P-V criticality in the extended phase space of Gauss-Bonnet black holes in AdS space,''
JHEP \textbf{09}, 005 (2013)
doi:10.1007/JHEP09(2013)005
[arXiv:1306.6233 [gr-qc]].
%407 citations counted in INSPIRE as of 07 Mar 2024

%\cite{Kastor:2009wy}
\bibitem{Kastor}
D.~Kastor, S.~Ray and J.~Traschen,
``Enthalpy and the Mechanics of AdS Black Holes,''
Class. Quant. Grav. \textbf{26}, 195011 (2009)
doi:10.1088/0264-9381/26/19/195011
[arXiv:0904.2765 [hep-th]].
%1093 citations counted in INSPIRE as of 07 Mar 2024

%\cite{Dolan:2010ha}
\bibitem{Dolan}
B.~P.~Dolan,
``The cosmological constant and the black hole equation of state,''
Class. Quant. Grav. \textbf{28}, 125020 (2011)
doi:10.1088/0264-9381/28/12/125020
[arXiv:1008.5023 [gr-qc]].
%554 citations counted in INSPIRE as of 07 Mar 2024

%\cite{Dolan:2011xt}
\bibitem{Dolan2}
B.~P.~Dolan,
``Pressure and volume in the first law of black hole thermodynamics,''
Class. Quant. Grav. \textbf{28}, 235017 (2011)
doi:10.1088/0264-9381/28/23/235017
[arXiv:1106.6260 [gr-qc]].
%574 citations counted in INSPIRE as of 07 Mar 2024

%\cite{Dolan3}
\bibitem{Dolan3}
B.~P.~Dolan,
``Compressibility of rotating black holes,''
Phys. Rev. D \textbf{84}, 127503 (2011)
doi:10.1103/PhysRevD.84.127503
[arXiv:1109.0198 [gr-qc]].
%189 citations counted in INSPIRE as of 07 Mar 2024


%\cite{Kubiznak:2016qmn}
\bibitem{Kubizna}
D.~Kubiznak, R.~B.~Mann and M.~Teo,
``Black hole chemistry: thermodynamics with Lambda,''
Class. Quant. Grav. \textbf{34}, no.6, 063001 (2017)
doi:10.1088/1361-6382/aa5c69
[arXiv:1608.06147 [hep-th]].
%577 citations counted in INSPIRE as of 07 Mar 2024

%\cite{Xu:2013zea}
\bibitem{Xu}
W.~Xu, H.~Xu and L.~Zhao,
``Gauss-Bonnet coupling constant as a free thermodynamical variable and the associated criticality,''
Eur. Phys. J. C \textbf{74}, 2970 (2014)
doi:10.1140/epjc/s10052-014-2970-8
[arXiv:1311.3053 [gr-qc]].
%113 citations counted in INSPIRE as of 07 Mar 2024

%\cite{Xu:2014kwa}
\bibitem{Xu2}
W.~Xu and L.~Zhao,
``Critical phenomena of static charged AdS black holes in conformal gravity,''
Phys. Lett. B \textbf{736}, 214-220 (2014)
doi:10.1016/j.physletb.2014.07.019
[arXiv:1405.7665 [gr-qc]].
%104 citations counted in INSPIRE as of 07 Mar 2024

%\cite{Zhang:2017lhl}
\bibitem{Zhang}
M.~Zhang, D.~C.~Zou and R.~H.~Yue,
``Reentrant phase transitions and triple points of topological AdS black holes in Born-Infeld-massive gravity,''
Adv. High Energy Phys. \textbf{2017}, 3819246 (2017)
doi:10.1155/2017/3819246
[arXiv:1707.04101 [hep-th]].
%38 citations counted in INSPIRE as of 07 Mar 2024


\bibitem{Ruppeiner:2012uc}
G.~Ruppeiner,
Thermodynamic curvature: pure fluids to black holes,
J. Phys. Conf. Ser. \textbf{410}, 012138 (2013)
doi:10.1088/1742-6596/410/1/012138
[arXiv:1210.2011 [gr-qc]].
%28 citations counted in INSPIRE as of 06 Mar 2024
%\cite{Miao:2017cyt}
\bibitem{Miao:2017cyt}
Y.~G.~Miao and Z.~M.~Xu,
Microscopic structures and thermal stability of black holes conformally coupled to scalar fields in five dimensions,
Nucl. Phys. B \textbf{942}, 205-220 (2019)
doi:10.1016/j.nuclphysb.2019.03.015
[arXiv:1711.01757 [hep-th]].
%41 citations counted in INSPIRE as of 07 Mar 2024
%\cite{Guo:2019oad}
\bibitem{Guo:2019oad}
X.~Y.~Guo, H.~F.~Li, L.~C.~Zhang and R.~Zhao,
Microstructure and continuous phase transition of a Reissner-Nordstrom-AdS black hole,
Phys. Rev. D \textbf{100}, no.6, 064036 (2019)
doi:10.1103/PhysRevD.100.064036
[arXiv:1901.04703 [gr-qc]].
%49 citations counted in INSPIRE as of 07 Mar 2024
%\cite{Wei:2019yvs}
\bibitem{Wei:2019yvs}
S.~W.~Wei, Y.~X.~Liu and R.~B.~Mann,
Ruppeiner Geometry, Phase Transitions, and the Microstructure of Charged AdS Black Holes,
Phys. Rev. D \textbf{100}, no.12, 124033 (2019)
doi:10.1103/PhysRevD.100.124033
[arXiv:1909.03887 [gr-qc]].
%101 citations counted in INSPIRE as of 07 Mar 2024
%\cite{Wang:2019cax}
\bibitem{Wang:2019cax}
P.~Wang, H.~Wu and H.~Yang,
Thermodynamic Geometry of AdS Black Holes and Black Holes in a Cavity,
Eur. Phys. J. C \textbf{80}, no.3, 216 (2020)
doi:10.1140/epjc/s10052-020-7776-2
[arXiv:1910.07874 [gr-qc]].
%37 citations counted in INSPIRE as of 07 Mar 2024
%\cite{Yerra:2020oph}
\bibitem{Yerra:2020oph}
P.~K.~Yerra and C.~Bhamidipati,
Ruppeiner Geometry, Phase Transitions and Microstructures of Black Holes in Massive Gravity,
Int. J. Mod. Phys. A \textbf{35}, no.22, 2050120 (2020)
doi:10.1142/S0217751X20501201
[arXiv:2006.07775 [hep-th]].
%33 citations counted in INSPIRE as of 07 Mar 2024
%\cite{Yerra:2021hnh}
\bibitem{Yerra:2021hnh}
P.~K.~Yerra and C.~Bhamidipati,
Novel relations in massive gravity at Hawking-Page transition,
Phys. Rev. D \textbf{104}, no.10, 104049 (2021)
doi:10.1103/PhysRevD.104.104049
[arXiv:2107.04504 [gr-qc]].
%13 citations counted in INSPIRE as of 07 Mar 2024


\bibitem{Wu:2022whe}
D.~Wu,
Topological classes of rotating black holes,
Phys. Rev. D \textbf{107}, no.2, 024024 (2023)
doi:10.1103/PhysRevD.107.024024
[arXiv:2211.15151 [gr-qc]].
%7 citations counted in INSPIRE as of 11 Apr 2023
%\cite{Liu:2022aqt}
\bibitem{Liu:2022aqt}
C.~Liu and J.~Wang,
Topological natures of the Gauss-Bonnet black hole in AdS space,
Phys. Rev. D \textbf{107}, no.6, 064023 (2023)
doi:10.1103/PhysRevD.107.064023
[arXiv:2211.05524 [gr-qc]].
%9 citations counted in INSPIRE as of 11 Apr 2023
%\cite{Fan:2022bsq}
\bibitem{Fan:2022bsq}
Z.~Y.~Fan,
Topological interpretation for phase transitions of black holes,
Phys. Rev. D \textbf{107}, no.4, 044026 (2023)
doi:10.1103/PhysRevD.107.044026
[arXiv:2211.12957 [gr-qc]].

\bibitem{Gogoi:2023xzy}
N.~J.~Gogoi and P.~Phukon,
Thermodynamic topology of 4D dyonic AdS black holes in different ensembles,
Phys. Rev. D \textbf{108}, no.6, 066016 (2023)
doi:10.1103/PhysRevD.108.066016
[arXiv:2304.05695 [hep-th]].



%1 citations counted in INSPIRE as of 10 Apr 2023
\bibitem{Ali:2023zww}
M.~S.~Ali, H.~El Moumni, J.~Khalloufi and K.~Masmar,
Topology of Born-Infeld-AdS Black Hole Phase Transition,
[arXiv:2306.11212 [hep-th]].
%1 citations counted in INSPIRE as of 30 Jun 2023

%0 citations counted in INSPIRE as of 14 Dec 2023
\bibitem{Saleem:2023oue}
M.~A.~Saleem and A.~Taani,
The chaotic behavior of black holes: Investigating a topological retraction in anti-de Sitter spaces,
New Astron. \textbf{107}, 102149 (2024)
doi:10.1016/j.newast.2023.102149
%0 citations counted in INSPIRE as of 14 Dec 2023
\bibitem{Shahzad:2023cis}
M.~U.~Shahzad, A.~Mehmood, S.~Sharif and A.~\"Ovg\"un,
Criticality and topological classes of neutral Gauss\textendash{}Bonnet AdS black holes in 5D,
Annals Phys. \textbf{458}, no.3, 169486 (2023)
doi:10.1016/j.aop.2023.169486
%2 citations counted in INSPIRE as of 14 Dec 2023
\bibitem{Chen:2023elp}
Z.~Q.~Chen and S.~W.~Wei,
Thermodynamics, Ruppeiner geometry, and topology of Born-Infeld black hole in asymptotic flat spacetime,
Nucl. Phys. B \textbf{996}, 116369 (2023)
doi:10.1016/j.nuclphysb.2023.116369
%1 citations counted in INSPIRE as of 14 Dec 2023
\bibitem{Bai:2022klw}
N.~C.~Bai, L.~Li and J.~Tao,
Topology of black hole thermodynamics in Lovelock gravity,
Phys. Rev. D \textbf{107}, no.6, 064015 (2023)
doi:10.1103/PhysRevD.107.064015
[arXiv:2208.10177 [gr-qc]].
%39 citations counted in INSPIRE as of 15 Dec 2023
\bibitem{Yerra:2022alz}
P.~K.~Yerra and C.~Bhamidipati,
Topology of black hole thermodynamics in Gauss-Bonnet gravity,
Phys. Rev. D \textbf{105}, no.10, 104053 (2022)
doi:10.1103/PhysRevD.105.104053
[arXiv:2202.10288 [gr-qc]].
%38 citations counted in INSPIRE as of 15 Dec 2023
%\cite{Hazarika:2023iwp}
\bibitem{Hazarika:2023iwp}
B.~Hazarika and P.~Phukon,
Thermodynamic Topology of $D=4,5$ Horava Lifshitz Black Hole in Two Ensembles,
[arXiv:2312.06324 [hep-th]].
%0 citations counted in INSPIRE as of 15 Dec 2023


\bibitem{Liu:2014gvf}
Y.~Liu, D.~C.~Zou and B.~Wang,
Signature of the Van der Waals like small-large charged AdS black hole phase transition in quasinormal modes,
JHEP \textbf{09}, 179 (2014)
doi:10.1007/JHEP09(2014)179
[arXiv:1405.2644 [hep-th]].
%109 citations counted in INSPIRE as of 06 Mar 2024
%\cite{Zou:2017juz}
\bibitem{Zou:2017juz}
D.~C.~Zou, Y.~Liu and R.~H.~Yue,
Behavior of quasinormal modes and Van der Waals-like phase transition of charged AdS black holes in massive gravity,
Eur. Phys. J. C \textbf{77}, no.6, 365 (2017)
doi:10.1140/epjc/s10052-017-4937-z
[arXiv:1702.08118 [gr-qc]].
%70 citations counted in INSPIRE as of 06 Mar 2024
%\cite{Zhang:2020khz}
\bibitem{Zhang:2020khz}
M.~Zhang, C.~M.~Zhang, D.~C.~Zou and R.~H.~Yue,
Phase transition and Quasinormal modes for Charged black holes in 4D Einstein-Gauss-Bonnet gravity,
Chin. Phys. C \textbf{45}, no.4, 045105 (2021)
doi:10.1088/1674-1137/abe19a
[arXiv:2009.03096 [hep-th]].
%11 citations counted in INSPIRE as of 06 Mar 2024
%\cite{Mahapatra:2016dae}
\bibitem{Mahapatra:2016dae}
S.~Mahapatra,
Thermodynamics, Phase Transition and Quasinormal modes with Weyl corrections,
JHEP \textbf{04}, 142 (2016)
doi:10.1007/JHEP04(2016)142
[arXiv:1602.03007 [hep-th]].
%44 citations counted in INSPIRE as of 06 Mar 2024
%\cite{Chabab:2016cem}
\bibitem{Chabab:2016cem}
M.~Chabab, H.~El Moumni, S.~Iraoui and K.~Masmar,
Behavior of quasinormal modes and high dimension RN\textendash{}AdS black hole phase transition,
Eur. Phys. J. C \textbf{76}, no.12, 676 (2016)
doi:10.1140/epjc/s10052-016-4518-6
[arXiv:1606.08524 [hep-th]].
%53 citations counted in INSPIRE as of 06 Mar 2024


\bibitem{Wei:2017mwc}
S.~W.~Wei and Y.~X.~Liu,
Photon orbits and thermodynamic phase transition of $d$-dimensional charged AdS black holes,
Phys. Rev. D \textbf{97}, no.10, 104027 (2018)
doi:10.1103/PhysRevD.97.104027
[arXiv:1711.01522 [gr-qc]].
%57 citations counted in INSPIRE as of 06 Mar 2024
%\cite{Wei:2018aqm}
\bibitem{Wei:2018aqm}
S.~W.~Wei, Y.~X.~Liu and Y.~Q.~Wang,
Probing the relationship between the null geodesics and thermodynamic phase transition for rotating Kerr-AdS black holes,
Phys. Rev. D \textbf{99}, no.4, 044013 (2019)
doi:10.1103/PhysRevD.99.044013
[arXiv:1807.03455 [gr-qc]].
%27 citations counted in INSPIRE as of 06 Mar 2024
%\cite{Zhang:2019tzi}
\bibitem{Zhang:2019tzi}
M.~Zhang, S.~Z.~Han, J.~Jiang and W.~B.~Liu,
%``Circular orbit of a test particle and phase transition of a black hole,''
Phys. Rev. D \textbf{99}, no.6, 065016 (2019)
doi:10.1103/PhysRevD.99.065016
[arXiv:1903.08293 [hep-th]].
%20 citations counted in INSPIRE as of 06 Mar 2024



\bibitem{Zhang:2019glo}
M.~Zhang and M.~Guo,
Can shadows reflect phase structures of black holes?,
Eur. Phys. J. C \textbf{80}, no.8, 790 (2020)
doi:10.1140/epjc/s10052-020-8389-5
[arXiv:1909.07033 [gr-qc]].
%64 citations counted in INSPIRE as of 06 Mar 2024
%\cite{Belhaj:2020nqy}
\bibitem{Belhaj:2020nqy}
A.~Belhaj, L.~Chakhchi, H.~El Moumni, J.~Khalloufi and K.~Masmar,
Thermal Image and Phase Transitions of Charged AdS Black Holes using Shadow Analysis,
Int. J. Mod. Phys. A \textbf{35}, no.27, 2050170 (2020)
doi:10.1142/S0217751X20501705
[arXiv:2005.05893 [gr-qc]].

\bibitem{lyp}
A.M Lyapunov,
The general problem of the stability of motion
Int. J. Control 55, 531 (1992).
doi:https://doi.org/10.1080/00207179208934253
\bibitem{lyp2}
M. Sandri,
Numerical calculation of lyapunov exponents,
Math. J. 6, 78 (1996).


%\cite{Sorokhaibam:2019qho}
\bibitem{syk}
N.~Sorokhaibam,
``Phase transition and chaos in charged SYK model,''
JHEP \textbf{07}, 055 (2020)
doi:10.1007/JHEP07(2020)055
[arXiv:1912.04326 [hep-th]].
%23 citations counted in INSPIRE as of 12 May 2025
%\cite{Davis:2022iqi}
\bibitem{syk2}
A.~Davis and Y.~Wang,
``Quantum chaos and phase transition in the Yukawa\textendash{}Sachdev-Ye-Kitaev model,''
Phys. Rev. B \textbf{107}, no.20, 205122 (2023)
doi:10.1103/PhysRevB.107.205122
[arXiv:2212.03265 [cond-mat.str-el]].
%6 citations counted in INSPIRE as of 12 May 2025
%\cite{Emary:2003zza}
\bibitem{Dicke}
C.~Emary and T.~Brandes,
``Chaos and the quantum phase transition in the Dicke model,''
Phys. Rev. E \textbf{67}, 066203 (2003)
doi:10.1103/PhysRevE.67.066203
[arXiv:cond-mat/0301273 [cond-mat]].
%270 citations counted in INSPIRE as of 12 May 2025
%\cite{Miritello:2008zd}
\bibitem{coscll}
G.~Miritello, A.~Pluchino and A.~Rapisarda,
``Phase Transitions and Chaos in Long-Range Models of Coupled Oscillators,''
EPL \textbf{85}, no.1, 10007 (2009)
doi:10.1209/0295-5075/85/10007
[arXiv:0807.1870 [cond-mat.stat-mech]].
%2 citations counted in INSPIRE as of 12 May 2025
%\cite{Heiss:1991zza}
\bibitem{finite}
W.~D.~Heiss and A.~L.~Sannino,
``Transitional regions of finite Fermi systems and quantum chaos,''
Phys. Rev. A \textbf{43}, 4159-4166 (1991)
doi:10.1103/PhysRevA.43.4159
%31 citations counted in INSPIRE as of 12 May 2025


%\cite{Sota:1995ms}
\bibitem{static}
Y.~Sota, S.~Suzuki and K.~i.~Maeda,
``Chaos in static axisymmetric space-times. 1: Vacuum case,''
Class. Quant. Grav. \textbf{13}, 1241-1260 (1996)
doi:10.1088/0264-9381/13/5/034
[arXiv:gr-qc/9505036 [gr-qc]].
%75 citations counted in INSPIRE as of 12 May 2025
%\cite{Sota:1996cv}
\bibitem{static2}
Y.~Sota, S.~Suzuki and K.~i.~Maeda,
``Chaos in static axisymmetric space-times. 2. Nonvacuum case,''
[arXiv:gr-qc/9610065 [gr-qc]].
%6 citations counted in INSPIRE as of 12 May 2025
%\cite{Kan:2021blg}
\bibitem{kerr}
N.~Kan and B.~Gwak,
``Bound on the Lyapunov exponent in Kerr-Newman black holes via a charged particle,''
Phys. Rev. D \textbf{105}, no.2, 026006 (2022)
doi:10.1103/PhysRevD.105.026006
[arXiv:2109.07341 [gr-qc]].
%31 citations counted in INSPIRE as of 12 May 2025
%\cite{Gwak:2022xje}
\bibitem{kerr2}
B.~Gwak, N.~Kan, B.~H.~Lee and H.~Lee,
``Violation of bound on chaos for charged probe in Kerr-Newman-AdS black hole,''
JHEP \textbf{09}, 026 (2022)
doi:10.1007/JHEP09(2022)026
[arXiv:2203.07298 [gr-qc]].
%30 citations counted in INSPIRE as of 12 May 2025
%\cite{Hanan:2006uf}
\bibitem{multi}
W.~Hanan and E.~Radu,
``Chaotic motion in multi-black hole spacetimes and holographic screens,''
Mod. Phys. Lett. A \textbf{22}, 399-406 (2007)
doi:10.1142/S0217732307022815
[arXiv:gr-qc/0610119 [gr-qc]].
%28 citations counted in INSPIRE as of 12 May 2025
%\cite{Lu:2018mpr}
\bibitem{qg}
F.~Lu, J.~Tao and P.~Wang,
``Minimal Length Effects on Chaotic Motion of Particles around Black Hole Horizon,''
JCAP \textbf{12}, 036 (2018)
doi:10.1088/1475-7516/2018/12/036
[arXiv:1811.02140 [gr-qc]].
%30 citations counted in INSPIRE as of 12 May 2025
%\cite{Guo:2020xnf}
\bibitem{qg2}
X.~Guo, K.~Liang, B.~Mu, P.~Wang and M.~Yang,
``Chaotic Motion around a Black Hole under Minimal Length Effects,''
Eur. Phys. J. C \textbf{80}, no.8, 745 (2020)
doi:10.1140/epjc/s10052-020-8335-6
[arXiv:2002.05894 [gr-qc]].
%11 citations counted in INSPIRE as of 12 May 2025

%\cite{Maldacena:2015waa}
\bibitem{mss}
J.~Maldacena, S.~H.~Shenker and D.~Stanford,
``A bound on chaos,''
JHEP \textbf{08}, 106 (2016)
doi:10.1007/JHEP08(2016)106
[arXiv:1503.01409 [hep-th]].
%2007 citations counted in INSPIRE as of 12 May 2025
%\cite{Hashimoto:2016dfz}
\bibitem{hori}
K.~Hashimoto and N.~Tanahashi,
``Universality in Chaos of Particle Motion near Black Hole Horizon,''
Phys. Rev. D \textbf{95}, no.2, 024007 (2017)
doi:10.1103/PhysRevD.95.024007
[arXiv:1610.06070 [hep-th]].
%118 citations counted in INSPIRE as of 12 May 2025
%\cite{Dalui:2018qqv}
\bibitem{hori2}
S.~Dalui, B.~R.~Majhi and P.~Mishra,
``Presence of horizon makes particle motion chaotic,''
Phys. Lett. B \textbf{788}, 486-493 (2019)
doi:10.1016/j.physletb.2018.11.050
[arXiv:1803.06527 [gr-qc]].
%97 citations counted in INSPIRE as of 12 May 2025

%\cite{Zhao:2018wkl}
\bibitem{vio}
Q.~Q.~Zhao, Y.~Z.~Li and H.~Lu,
``Static Equilibria of Charged Particles Around Charged Black Holes: Chaos Bound and Its Violations,''
Phys. Rev. D \textbf{98}, no.12, 124001 (2018)
doi:10.1103/PhysRevD.98.124001
[arXiv:1809.04616 [gr-qc]].
%50 citations counted in INSPIRE as of 12 May 2025
%\cite{Guo:2020pgq}
\bibitem{vio2}
X.~Guo, K.~Liang, B.~Mu, P.~Wang and M.~Yang,
``Minimal Length Effects on Motion of a Particle in Rindler Space,''
Chin. Phys. C \textbf{45}, no.2, 023115 (2021)
doi:10.1088/1674-1137/abcf20
[arXiv:2007.07744 [gr-qc]].
%18 citations counted in INSPIRE as of 12 May 2025
%\cite{Gwak:2022xje}
\bibitem{vio3}
B.~Gwak, N.~Kan, B.~H.~Lee and H.~Lee,
``Violation of bound on chaos for charged probe in Kerr-Newman-AdS black hole,''
JHEP \textbf{09}, 026 (2022)
doi:10.1007/JHEP09(2022)026
[arXiv:2203.07298 [gr-qc]].
%30 citations counted in INSPIRE as of 12 May 2025
%\cite{Park:2023lfc}
\bibitem{vio4}
J.~Park and B.~Gwak,
``Bound on Lyapunov exponent in Kerr-Newman-de Sitter black holes by a charged particle,''
JHEP \textbf{04}, 023 (2024)
doi:10.1007/JHEP04(2024)023
[arXiv:2312.13075 [gr-qc]].
%3 citations counted in INSPIRE as of 12 May 2025
%\cite{Guo:2022kio}
\bibitem{first}
X.~Guo, Y.~Lu, B.~Mu and P.~Wang,
``Probing phase structure of black holes with Lyapunov exponents,''
JHEP \textbf{08}, 153 (2022)
doi:10.1007/JHEP08(2022)153
[arXiv:2205.02122 [gr-qc]].
%18 citations counted in INSPIRE as of 12 May 2025
%\cite{Yang:2023hci}
\bibitem{le}
S.~Yang, J.~Tao, B.~Mu and A.~He,
``Lyapunov exponents and phase transitions of Born-Infeld AdS black holes,''
JCAP \textbf{07}, 045 (2023)
doi:10.1088/1475-7516/2023/07/045
[arXiv:2304.01877 [gr-qc]].
%15 citations counted in INSPIRE as of 12 May 2025
%\cite{Lyu:2023sih}
\bibitem{le2}
X.~Lyu, J.~Tao and P.~Wang,
``Probing the thermodynamics of charged Gauss Bonnet AdS black holes with the Lyapunov exponent,''
Eur. Phys. J. C \textbf{84}, no.9, 974 (2024)
doi:10.1140/epjc/s10052-024-13354-9
[arXiv:2312.11912 [gr-qc]].
%10 citations counted in INSPIRE as of 12 May 2025
%\cite{Kumara:2024obd}
\bibitem{le3}
A.~N.~Kumara, S.~Punacha and M.~S.~Ali,
``Lyapunov exponents and phase structure of Lifshitz and hyperscaling violating black holes,''
JCAP \textbf{07}, 061 (2024)
doi:10.1088/1475-7516/2024/07/061
[arXiv:2401.05181 [gr-qc]].
%12 citations counted in INSPIRE as of 12 May 2025
%\cite{Du:2024uhd}
\bibitem{le4}
Y.~Z.~Du, H.~F.~Li, Y.~B.~Ma and Q.~Gu,
``Phase structure and optical properties of the de Sitter Spacetime with KR field based on the Lyapunov exponent,''
Eur. Phys. J. C \textbf{85}, no.1, 78 (2025)
doi:10.1140/epjc/s10052-025-13809-7
[arXiv:2403.20083 [hep-th]].
%12 citations counted in INSPIRE as of 12 May 2025
%\cite{Gogoi:2024akv}
\bibitem{le5}
N.~J.~Gogoi, S.~Acharjee and P.~Phukon,
``Lyapunov exponents and phase transition of Hayward AdS black hole,''
Eur. Phys. J. C \textbf{84}, no.11, 1144 (2024)
doi:10.1140/epjc/s10052-024-13520-z
[arXiv:2404.03947 [hep-th]].
%4 citations counted in INSPIRE as of 12 May 2025
%\cite{Shukla:2024tkw}
\bibitem{le6}
B.~Shukla, P.~P.~Das, D.~Dudal and S.~Mahapatra,
``Interplay between the Lyapunov exponents and phase transitions of charged AdS black holes,''
Phys. Rev. D \textbf{110}, no.2, 024068 (2024)
doi:10.1103/PhysRevD.110.024068
[arXiv:2404.02095 [hep-th]].
%6 citations counted in INSPIRE as of 12 May 2025
%\cite{Chen:2025xqc}
\bibitem{le7}
D.~Chen, C.~Yang and Y.~Liu,
``Lyapunov exponents as probes for a phase transition of a Kerr-AdS black hole,''
Phys. Lett. B \textbf{865}, 139463 (2025)
doi:10.1016/j.physletb.2025.139463
[arXiv:2501.16999 [hep-th]].
%1 citations counted in INSPIRE as of 12 May 2025
%\cite{R:2025gok}
\bibitem{le8}
K.~R., D.~D., K.~M.~Ajith, K.~Hegde, S.~Punacha and A.~N.~Kumara,
``Euclidean Thermodynamics and Lyapunov Exponents of Einstein-Power-Yang-Mills AdS Black Holes,''
[arXiv:2504.12890 [gr-qc]].
%0 citations counted in INSPIRE as of 12 May 2025


%\cite{Sahay:2010yq}
\bibitem{r1}
A.~Sahay, T.~Sarkar and G.~Sengupta,
``On The Phase Structure and Thermodynamic Geometry of R-Charged Black Holes,''
JHEP \textbf{11}, 125 (2010)
doi:10.1007/JHEP11(2010)125
[arXiv:1009.2236 [hep-th]].
%36 citations counted in INSPIRE as of 13 May 2025
%\cite{Gubser:2004xx}
\bibitem{r2}
S.~S.~Gubser and J.~J.~Heckman,
``Thermodynamics of R-charged black holes in AdS(5) from effective strings,''
JHEP \textbf{11}, 052 (2004)
doi:10.1088/1126-6708/2004/11/052
[arXiv:hep-th/0411001 [hep-th]].
%36 citations counted in INSPIRE as of 13 May 2025
%\cite{Jain:2009uj}
\bibitem{r3}
S.~Jain, S.~Mukherji and S.~Mukhopadhyay,
``Notes on R-charged black holes near criticality and gauge theory,''
JHEP \textbf{11}, 051 (2009)
doi:10.1088/1126-6708/2009/11/051
[arXiv:0906.5134 [hep-th]].
%21 citations counted in INSPIRE as of 13 May 2025
%\cite{Wang:2021llu}
\bibitem{r4}
P.~Wang and F.~Yao,
``Thermodynamic geometry of black holes enclosed by a cavity in extended phase space,''
Nucl. Phys. B \textbf{976}, 115715 (2022)
doi:10.1016/j.nuclphysb.2022.115715
[arXiv:2107.14640 [gr-qc]].
%11 citations counted in INSPIRE as of 13 May 2025
%\cite{Gogoi:2023qku}
\bibitem{r5}
N.~J.~Gogoi and P.~Phukon,
``Topology of thermodynamics in R-charged black holes,''
Phys. Rev. D \textbf{107}, no.10, 106009 (2023)
doi:10.1103/PhysRevD.107.106009
%43 citations counted in INSPIRE as of 13 May 2025

%\cite{Cvetic:1999ne}
\bibitem{rh}
M.~Cvetic and S.~S.~Gubser,
``Phases of R charged black holes, spinning branes and strongly coupled gauge theories,''
JHEP \textbf{04}, 024 (1999)
doi:10.1088/1126-6708/1999/04/024
[arXiv:hep-th/9902195 [hep-th]].
%475 citations counted in INSPIRE as of 13 May 2025
%\cite{Phukon:2013tda}
\bibitem{pjp}
P.~Phukon and T.~Sarkar,
``R-Charged Black Holes and Holographic Optics,''
JHEP \textbf{09}, 102 (2013)
doi:10.1007/JHEP09(2013)102
[arXiv:1305.2745 [hep-th]].
%8 citations counted in INSPIRE as of 13 May 2025
%\cite{Mahapatra:2013vta}
\bibitem{pjp2}
S.~Mahapatra, P.~Phukon and T.~Sarkar,
``Generalized Superconductors and Holographic Optics,''
JHEP \textbf{01}, 135 (2014)
doi:10.1007/JHEP01(2014)135
[arXiv:1305.6273 [hep-th]].
%15 citations counted in INSPIRE as of 13 May 2025
%\cite{Mahapatra:2014qfa}
\bibitem{Mahapatra}
S.~Mahapatra,
``Generalized superconductors and holographic optics. Part II,''
JHEP \textbf{01}, 148 (2015)
doi:10.1007/JHEP01(2015)148
[arXiv:1411.6405 [hep-th]].
%7 citations counted in INSPIRE as of 13 May 2025

%\cite{Son:2006em}
\bibitem{hydror}
D.~T.~Son and A.~O.~Starinets,
``Hydrodynamics of r-charged black holes,''
JHEP \textbf{03}, 052 (2006)
doi:10.1088/1126-6708/2006/03/052
[arXiv:hep-th/0601157 [hep-th]].
%270 citations counted in INSPIRE as of 13 May 2025
%\cite{Mas:2006dy}
\bibitem{shear}
J.~Mas,
``Shear viscosity from R-charged AdS black holes,''
JHEP \textbf{03}, 016 (2006)
doi:10.1088/1126-6708/2006/03/016
[arXiv:hep-th/0601144 [hep-th]].
%111 citations counted in INSPIRE as of 13 May 2025

%\cite{Duff:1999gh}
\bibitem{Duff}
M.~J.~Duff and J.~T.~Liu,
``Anti-de Sitter black holes in gauged N = 8 supergravity,''
Nucl. Phys. B \textbf{554}, 237-253 (1999)
doi:10.1016/S0550-3213(99)00299-0
[arXiv:hep-th/9901149 [hep-th]].
%277 citations counted in INSPIRE as of 14 May 2025


\bibitem{0th}
V.P. Maslov
``Zeroth-Order Phase Transitions"
Mathematical Notes \textbf{76}, 697–710 (2004)
https://doi.org/10.1023/B:MATN.0000049669.32515.f0
%\cite{Gunasekaran:2012dq}
\bibitem{0th2}
S.~Gunasekaran, R.~B.~Mann and D.~Kubiznak,
``Extended phase space thermodynamics for charged and rotating black holes and Born-Infeld vacuum polarization,''
JHEP \textbf{11}, 110 (2012)
doi:10.1007/JHEP11(2012)110
[arXiv:1208.6251 [hep-th]].
%644 citations counted in INSPIRE as of 15 May 2025
%\cite{Altamirano:2013ane}
\bibitem{0th3}
N.~Altamirano, D.~Kubiznak and R.~B.~Mann,
``Reentrant phase transitions in rotating anti\textendash{}de Sitter black holes,''
Phys. Rev. D \textbf{88}, no.10, 101502 (2013)
doi:10.1103/PhysRevD.88.101502
[arXiv:1306.5756 [hep-th]].
%396 citations counted in INSPIRE as of 15 May 2025
%\cite{Hennigar:2015wxa}
\bibitem{0th4}
R.~A.~Hennigar and R.~B.~Mann,
``Reentrant phase transitions and van der Waals behaviour for hairy black holes,''
Entropy \textbf{17}, no.12, 8056-8072 (2015)
doi:10.3390/e17127862
[arXiv:1509.06798 [hep-th]].
%105 citations counted in INSPIRE as of 15 May 2025

%\cite{Dehyadegari:2017flm}
\bibitem{0th5}
A.~Dehyadegari, A.~Sheykhi and A.~Montakhab,
``Novel phase transition in charged dilaton black holes,''
Phys. Rev. D \textbf{96}, no.8, 084012 (2017)
doi:10.1103/PhysRevD.96.084012
[arXiv:1707.05307 [hep-th]].
%41 citations counted in INSPIRE as of 15 May 2025
%\cite{Dayyani:2017fuz}
\bibitem{0th6}
Z.~Dayyani, A.~Sheykhi, M.~H.~Dehghani and S.~Hajkhalili,
``Critical behavior and phase transition of dilaton black holes with nonlinear electrodynamics,''
Eur. Phys. J. C \textbf{78}, no.2, 152 (2018)
doi:10.1140/epjc/s10052-018-5623-5
[arXiv:1709.06875 [gr-qc]].
%45 citations counted in INSPIRE as of 15 May 2025
%\cite{Chen:2018icg}
\bibitem{0th7}
Y.~Chen, H.~Li and S.~J.~Zhang,
``Microscopic explanation for black hole phase transitions via Ruppeiner geometry: Two competing factors\textendash{}the temperature and repulsive interaction among BH molecules,''
Nucl. Phys. B \textbf{948}, 114752 (2019)
doi:10.1016/j.nuclphysb.2019.114752
[arXiv:1812.11765 [hep-th]].
%21 citations counted in INSPIRE as of 15 May 2025

%\cite{Banerjee:2012zm}
\bibitem{ce}
R.~Banerjee and D.~Roychowdhury,
``Critical behavior of Born Infeld AdS black holes in higher dimensions,''
Phys. Rev. D \textbf{85}, 104043 (2012)
doi:10.1103/PhysRevD.85.104043
[arXiv:1203.0118 [gr-qc]].
%116 citations counted in INSPIRE as of 21 May 2025






\end{thebibliography}
\end{document}